\documentclass[prx, english, twocolumn, longbibliography]{revtex4-1} 
\usepackage{bm, amsmath, amsfonts, amssymb, braket,mathtools,upgreek,amsthm}
\DeclareMathOperator{\Tr}{Tr}
\DeclareMathOperator{\im}{im}
\usepackage{multirow}
\usepackage[dvipdfmx]{graphicx}
\usepackage{float}
\usepackage{color}
\usepackage{comment}
\usepackage{subfigure}
\usepackage{lipsum}

\newtheorem{theorem}{Theorem}

\begin{document}

\newcommand{\ii}{\text{i}}
\newcommand{\U}{U}
\newcommand{\V}{V}
\newcommand{\BZ}{\left[ 0, 2\pi \right]}
\newcommand{\CZ}{\left[ 0, 1 \right]}

\title{Homotopy invariant in time-reversal and twofold rotation symmetric systems}

\author{Haoshu Li}
    \email{lihaoshu@mail.ustc.edu.cn}
    \affiliation{Department of Modern Physics, University of Science and Technology of China, Hefei 230026, China}

\author{Shaolong Wan}
    \email{slwan@ustc.edu.cn}

    \affiliation{Department of Modern Physics, University of Science and Technology of China, Hefei 230026, China}

\begin{abstract}
    The primary goal of this paper is to study topological invariants in two dimensional twofold rotation and time-reversal symmetric spinful systems. In this paper, firstly we build a new homotopy invariant based on the lifting of the Wilson loop to the universal covering group of the special orthogonal group. Furthermore, we prove that the invariant we built agrees with the K theory invariant. We go beyond the previous understanding of the Wilson loop unwinding in more than two occupied bands by finding an obstruction of such unwinding. Then, within this formalism, we show two examples that have the same Wilson loop spectrum but belong to different topological classes. Finally, we present a tight binding model realizing the non-trivial phase.
\end{abstract}

\maketitle

\section{Introduction}
    \label{sec: introduction}
Over the past few decades, there have been many studies on topological phases beyond the Landau paradigm \cite{PhysRevLett.45.494,PhysRevLett.48.1559,PhysRevB.23.5632,PhysRevB.31.3372} in particular on the subject of topological insulators and topological superconductors \cite{kane2005quantum,Bernevig1757,PhysRevLett.95.146802,PhysRevLett.98.106803,PhysRevB.74.195312,PhysRevB.78.195424}. Symmetries play an important role in researches. The internal symmetries implemented in fermionic systems are summarized by the 10 Altland-Zirnbauer(AZ) symmetry classes \cite{PhysRevB.55.1142,doi:10.1063/1.531675}. The classification of non-interacting topological phases in 10 AZ classes was systematically achieved by the K theory method \cite{doi:10.1063/1.3149495,PhysRevB.82.115120}. In the past few years, topological insulators and superconductors that are protected not only by internal symmetries but also by crystalline symmetries have been studied intensively. Topological insulators protected by crystalline symmetries are called topological crystalline insulators (TCIs) \cite{PhysRevLett.106.106802}. TCIs give rise to interesting new features, such as the presence of gapless surface Dirac cones pinned to mirror planes \cite{Hsieh2012,PhysRevB.90.081112}, and high-order topological insulators (HOTIs) featuring corner charges, corner states or hinge states \cite{PhysRevB.98.081110,PhysRevB.96.245115,PhysRevLett.119.246402,PhysRevB.99.245151}. In the study of TCIs, one important goal is finding topological invariants protected by symmetries. Many methods other than the Berry curvature method have been discovered to formulate topological invariants, for instance, symmetry indicators at high symmetry points \cite{PhysRevX.7.041069,PhysRevB.83.245132,PhysRevB.86.115112,PhysRevB.99.245151,Po2017}, Wilson loop methods \cite{PhysRevB.93.205104,PhysRevX.6.021008,PhysRevLett.121.106403,PhysRevB.100.115160,Hsieh2012,PhysRevB.96.245115} and elementary band representations \cite{Po2017,Bradlyn2017}.

The focus of this paper is to survey two dimensional time-reversal(TR) and twofold rotation symmetric spinful systems. The K theory classification of this system was completed \cite{PhysRevB.90.165114} and the result implies the existence of a new $\mathbb{Z}_2$ topological invariant. This topological invariant is described by vortices configuration at the high symmetry point in Ref. \cite{PhysRevB.100.075116}. However, to generate a vortex configuration, a smooth gauge is needed that is usually hard to implement. This system was also studied in Ref. \cite{PhysRevB.100.115160}, but the topological invariant described in that paper is more focused on Wannier band topology of the system and its obstructed atomic insulator nature, and the Wilson loop spectrum considered there is gapped. In this paper, we will show a phase with a non-trivial new topological invariant having a gapless Wilson loop spectrum instead. Until now, there has been a gap between the formulation of the new topological invariant and the K theory classification.

In this paper, we develop a new topological invariant using homotopy theory, and we prove that it agrees with the K theory invariant. We build this new homotopy invariant based on the lifting of the Wilson loop to the universal covering group. We will show in our formulation that the topology origin of the new $\mathbb{Z}_2$ topological invariant is disconnectedness of some fixed point set, hence its meaning is transparent. We further show that while nontrivial Wilson loop winding in the space of two occupied bands is protected by symmetries, and that Wilson loop unwinding can occur when embedded in higher-dimensional band space (as discussed in Refs. \cite{PhysRevB.99.045140,PhysRevB.100.195135}), there is an additional obstruction in a particular case of four occupied bands whose existence prevents the unwinding of the Wilson loop spectrum. This is characterized by the new $\mathbb{Z}_2$ topological invariant presented herein.

The paper is organized as follows: In Sec. \ref{sec: classification_review}, we~review the topological classification of two dimensional twofold rotation and TR-symmetric systems and three known topological invariants of these systems. In addition, we present a picture of what the new $\mathbb{Z}_2$ topological invariant is describing. In Sec. \ref{sec: Theory}, we build our theory on cases of two and four occupied bands. In Sec. \ref{sec: invariant}, a new homotopy invariant is formulated, and we provide proof that this homotopy invariant agrees with the K theory invariant. Then we show in Sec. \ref{sec: wil_examples} two examples distinguished by two symmetry categories that have the same Wilson loop spectrum but belong to different topological classes. In Sec. \ref{sec: ham_Example}, we introduce a Hamiltonian that analytically implements the Wilson loops in the previous sections, we present some numerical results on its tight binding model, and we discuss the physical meaning of the new topological invariant. Finally, we give our conclusions in Sec. \ref{sec: conclusion}.

\section{Topological classification of $T$ and $C_2$ symmetric system} \label{sec: classification_review}
In this paper, we focus on twofold rotation symmetric two dimensional systems in AZ class AII. The classification of these systems has been obtained in Ref. \cite{PhysRevB.90.165114}, since all symmetries here are order 2 symmetries. The system has two $\mathbb{Z}_2$ strong indices and two $\mathbb{Z}_2$ weak indices. The two $\mathbb{Z}_2$ weak indices are partial polarizations introduced by Fu and Kane along the $x$- and $y$-directions respectively \cite{PhysRevB.74.195312,PhysRevB.94.165164,PhysRevB.100.115160,PhysRevB.102.085108}, which are expressed as
\begin{align}
    \nu_{\Gamma X} = & \frac{1}{\pi} \left[ \int^{\pi}_{0} d k_{x} \Tr \mathcal{A}_x(k_x,0) + i \ln \frac{Pf[ w(\pi,0)]}{Pf[ w(0,0)]} \right] \notag \\
    & \mod 2, \notag \\
    \nu_{\Gamma Y} = & \frac{1}{\pi} \left[ \int^{\pi}_{0} d k_{y} \Tr \mathcal{A}_y(0,k_y) + i \ln \frac{Pf[ w(0,\pi)]}{Pf[ w(0,0)]} \right] \notag \\
    & \mod 2,
\end{align}
where $w_{mn}(k_x,k_y) = \bra{u_m(-k_x,-k_y)}T\ket{u_n(k_x,k_y)}$ is the sewing matrix of the TR-symmetry, $u_m(k_x,k_y)$ is the periodic part of the Bloch wave function, and $\mathcal{A}_x(k_x,k_y)$($\mathcal{A}_y(k_x,k_y)$) are Berry connections along the $x$($y$)-direction. Note that these two partial polarizations are quantized by the twofold rotation symmetry. As for the two $\mathbb{Z}_2$ strong indices, one of the two $\mathbb{Z}_2$ strong indices is Fu-Kane-Mele invariant in time-reversal invariant insulators \cite{kane2005quantum, PhysRevLett.95.146802, PhysRevB.74.195312}. Another one has been suggested to be characterized by vortices at high symmetry points \cite{PhysRevB.100.075116}, the topological invariant defined at each $C_2$-symmetric channel \cite{PhysRevB.102.041122,Kooi2021}. However, topological invariants defined in this way have not been shown directly related to the K theory classification and their $\mathbb{Z}_2$ topological nature is not transparent. We adopt a homotopic method to study this new $\mathbb{Z}_2$ strong index in this paper. For convenience, we simply denote this new $\mathbb{Z}_2$ strong topological invariant as $\nu_{new}$.

\begin{figure}
    \centering
    \subfigure[]{
        \centering
        \includegraphics[width=.3\columnwidth]{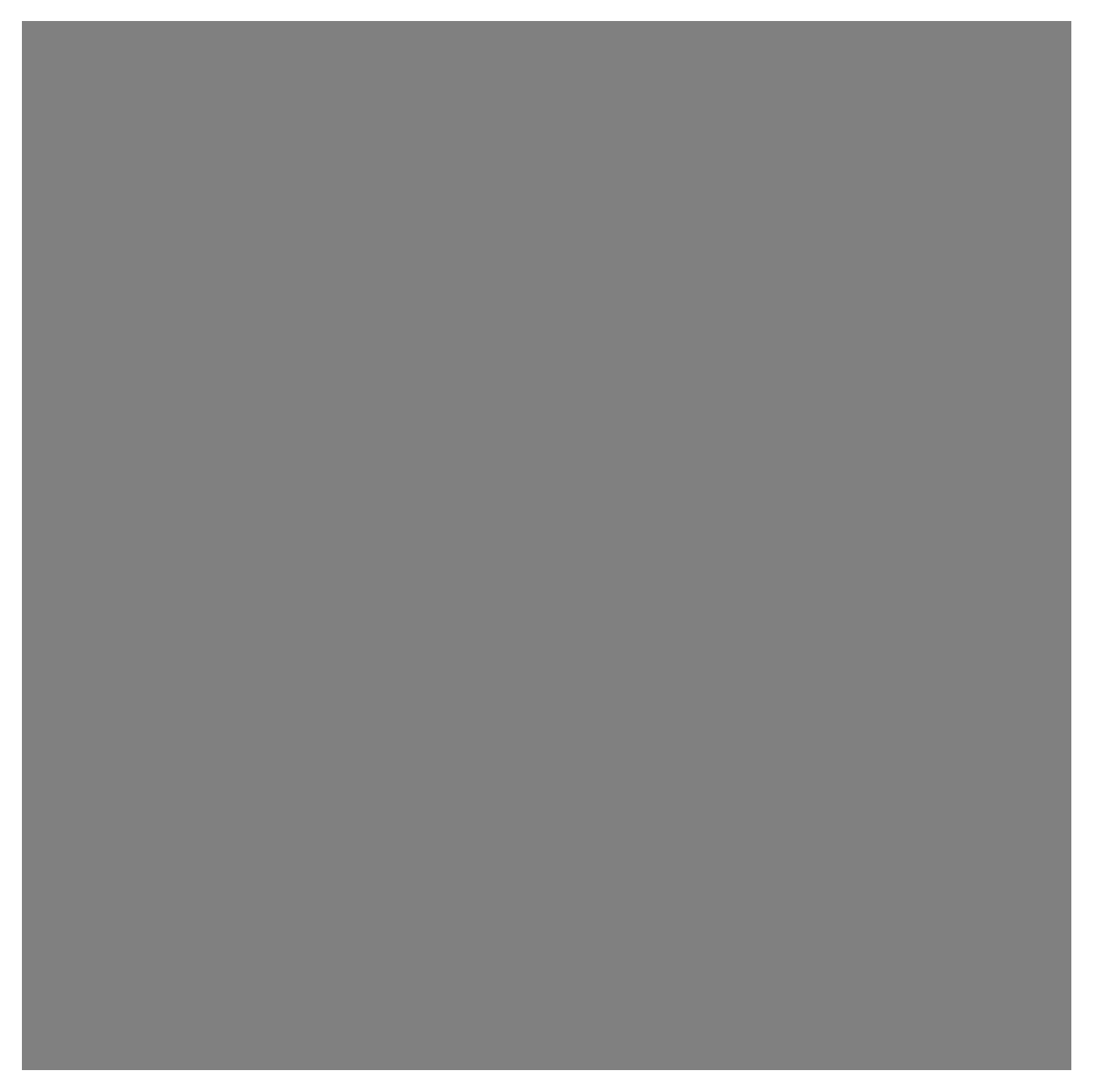}
    } 
    \subfigure[]{
        \centering
        \includegraphics[width=.3\columnwidth]{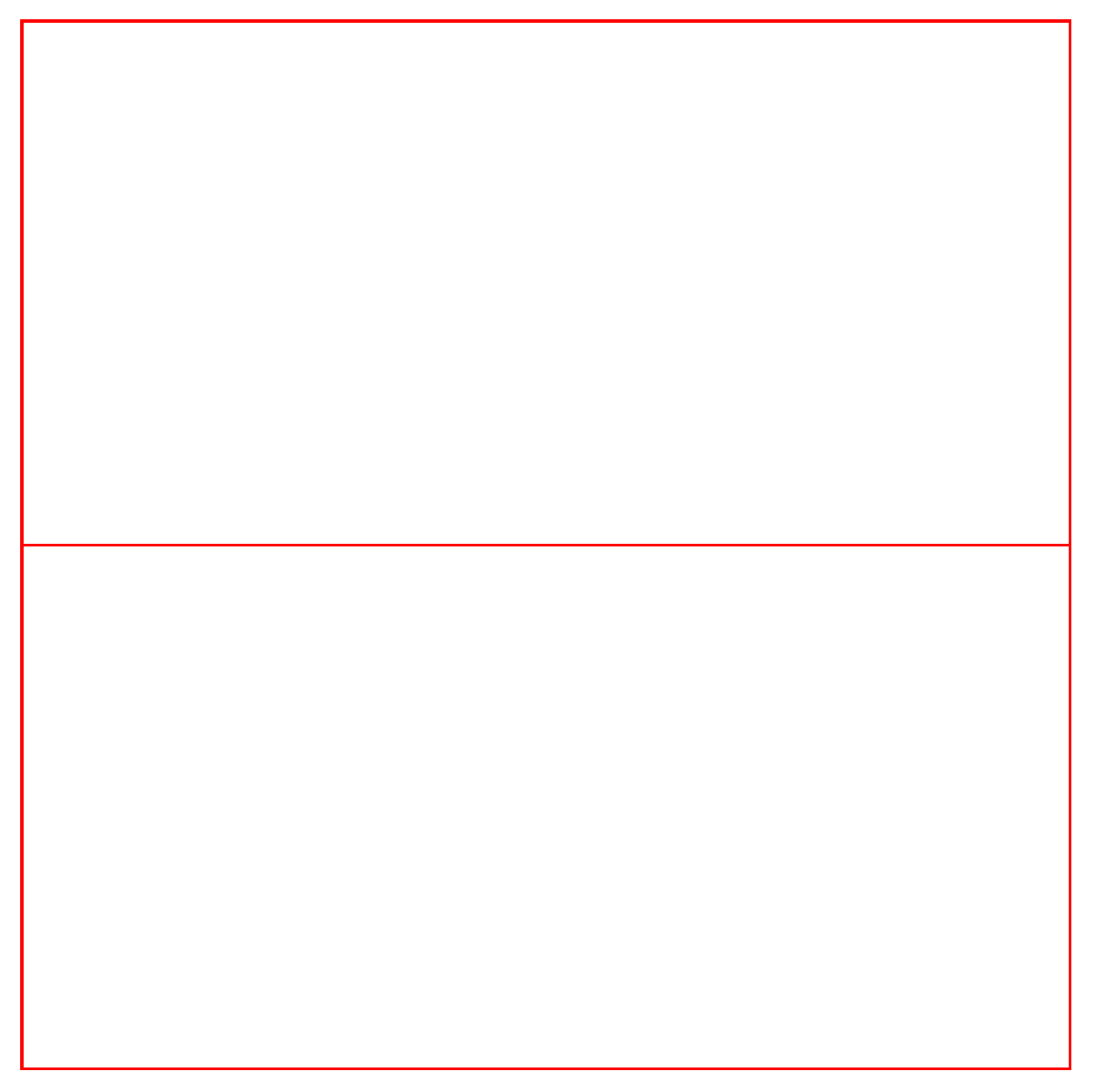}
    }
    \subfigure[]{
        \centering
        \includegraphics[width=.3\columnwidth]{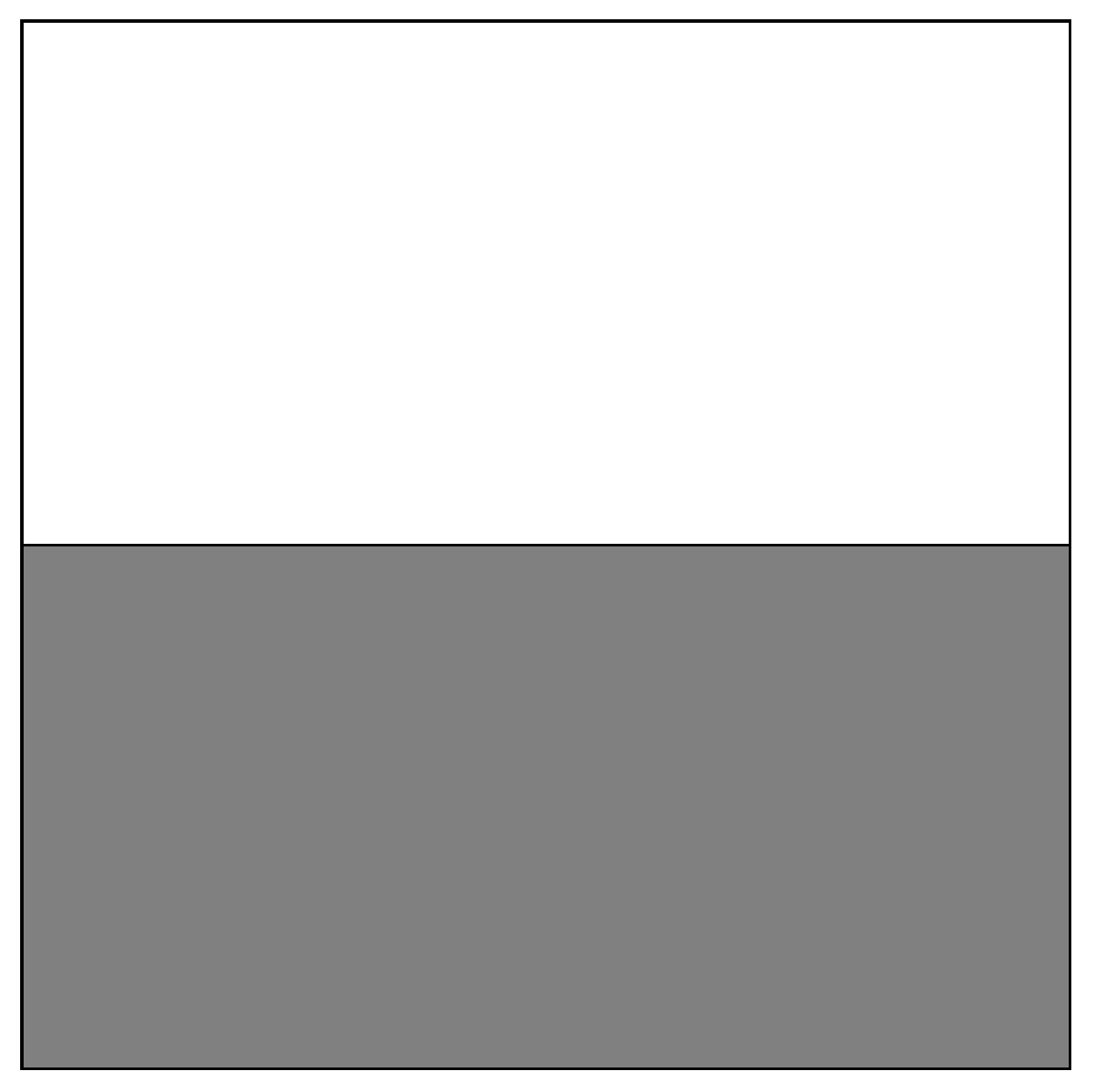}
    } 
    \caption[Cell decomposition of the Brillouin zone]{(a) Two dimensional skeleton $X_2$ is shown, which is the whole Brillouin zone. (b) One dimensional skeleton $X_1$ is shown in red color. When $\nu_{\Gamma X}= \nu_{\Gamma Y}= \nu_{\mathrm{FKM}}=0$, the Hamiltonian of the system can be continuously deformed to a Hamiltonian which is constant on $X_1$. (c) The effective half Brillouin zone is shown in gray color, which can further collapse to a sphere(the boundary of the effective half Brillouin zone collapses to a point) corresponding to one of two components of the space $X_2/X_1$. } \label{fig: Cell}
\end{figure}

Now, we introduce the concept and notation of Ref. \cite{Freed2013,PhysRevB.95.235425}. For a system with space-group action $G$, the space-group action on a Hamiltonian has a ``twist'' $(\tau,c)$, where $\tau$ is the factor system of the symmetry group and $c(g)=1(-1)$ indicates that the symmetry element $g$ is a symmetry(anti-symmetry). Antiunitary symmetries are specified by a $\mathbb{Z}_2$-valued function $\phi$ for group elements. Then an abelian group $\prescript{\phi}{}{K}^{(\tau,c)}_G(X)$ which characterizes the classification of the system can be introduced. The K group $\prescript{\phi}{}{K}^{(\tau,c)}_G(\mathbb{T}^2)$ for the Brillouin zone torus $\mathbb{T}^2$ provides a topological classification of two dimensional crystalline insulators subject to symmetry group $G$.

Make a cell decomposition with respect to symmetries as in Figure \ref{fig: Cell}. We give in appendix \ref{sec: K group} that the remaining new $\mathbb{Z}_2$ strong index $\nu_{new}$ corresponds to the relative K group $\prescript{\phi}{}{K}^{(\tau,c)}_G(X_2,X_1)$ in the framework of twisted equivariant K theory \cite{10.1093/qmath/17.1.367,Freed2013,PhysRevB.95.235425}. The classification K group of the system is
\begin{align}
    \prescript{\phi}{}{K}^{(\tau,c)}_G(\mathbb{T}^2) \cong \mathbb{Z} \oplus \mathbb{Z}_2^3 \oplus \prescript{\phi}{}{K}^{(\tau,c)}_G(X_2,X_1),
\end{align}
where the first $\mathbb{Z}$ summand is characterized by the number of occupied bands of the system(an even integer), and the second three $\mathbb{Z}_2$ summands are characterized by three topological invariants $\nu_{\Gamma X},\nu_{\Gamma Y},\nu_{\mathrm{FKM}}$, as we have mentioned. Hence, topological invariants $\nu_{\Gamma X},\nu_{\Gamma Y},\nu_{\mathrm{FKM}}$ are topological obstruction invariants without which the reduced K group is only expressed in terms of the relative K group $\prescript{\phi}{}{K}^{(\tau,c)}_G(X_2,X_1)$.

We also point out here that when the system satisfies $\nu_{\Gamma X}= \nu_{\Gamma Y}= \nu_{\mathrm{FKM}}=0$, since $\prescript{\phi}{}{K}^{(\tau,c)}_G(X_2,X_1) \cong \prescript{\phi}{}{\tilde{K}}^{(\tau,c)}_{\mathbb{Z}_2^{T C_2}}((X_2/X_1)_1) \cong \widetilde{KO}(S^2) \cong \mathbb{Z}_2$ where $(X_2/X_1)_1$ denotes one of the two components of $X_2/X_1$, the Hamiltonian of the system can be continuously deformed \footnote{Actually the Hamiltonian is continuously deformable in stable sense, i.e. a topological trivial Hamiltonian can be added in the deformation process.} to a Hamiltonian whose value is constant on the boundary of the effective half Brillouin zone $BZ_{\frac{1}{2}}$. Then we can view the whole system as two Hamiltonians over two disjoint two dimensional spheres which are time reversal related. The other three topological invariants $\nu_{\Gamma X},\nu_{\Gamma Y},\nu_{\mathrm{FKM}}$ are topological obstructions of such deformation. The new $\mathbb{Z}_2$ strong index is characterized by the second Stiefel-Whitney number of Hamiltonian over the effective half Brillouin zone $BZ_{\frac{1}{2}}$ since the K group element in $\prescript{\phi}{}{\tilde{K}}^{(\tau,c)}_{\mathbb{Z}_2^{T C_2}}((X_2/X_1)_1) \cong \widetilde{KO}(S^2)$ is captured by the second Stiefel-Whitney number over the sphere \cite{PhysRevLett.121.106403,10.2307/j.ctt1b7x751}, which can be read from the Wilson loop spectrum \cite{PhysRevLett.121.106403}. We call the second Stiefel-Whitney number over the effective half Brillouin zone the K theory invariant. However, we can not compute the new topological invariant $\nu_{new}$ in this way in general as the deformation of the Hamiltonian is not easy to find. In this paper, we will develop a general Wilson loop method to compute $\nu_{new}$ and prove that it agrees with the K theory invariant.

Now, we review how to read the other three topological invariants  $\nu_{\Gamma X},\nu_{\Gamma Y},\nu_{\mathrm{FKM}}$ from the Wilson loop spectrum. We first make a notation convention that is used throughout this paper. Denote the Wilson loop operator $W_{(k_x,k_y+2 \pi) \leftarrow (k_x,k_y)}$ used in Ref. \cite{PhysRevB.96.245115} by $W_{\mathbf{e}_2,k_y}(k_x)$. The path is a straight line from the starting point $(k_x,k_y)$ to the end point $(k_x,k_y+2 \pi)$ along the $k_y$-direction. The Wannier centers $v^y_j(k_x)$ and $v^x_j(k_y)$, where $k_x,k_y \in [-\pi,\pi]$, can be read from the Wilson loop spectrum with respect to any chosen branch, say $(-\frac{1}{2},\frac{1}{2}]$. Then,
\begin{align}
    \nu_{\Gamma X} & = \sum_j v^x_j(k_y=0) \mod 2 \notag ,\\
    \nu_{\Gamma Y} & = \sum_j v^y_j(k_x=0) \mod 2 \notag ,\\
    \nu_{\mathrm{FKM}} & = \sum_j [ v^x_j(k_y=0) + v^x_j(k_y=\pi) ] \mod 2\notag \\
    & = \sum_j [ v^y_j(k_x=0) + v^y_j(k_x=\pi) ] \mod 2.
\end{align}
The Wannier centers of a $T, C_2$-symmetric system satisfy symmetry constraints
\begin{align}
    \{ v_j(k_x) \} \stackrel{T}{=} \{ v_j(-k_x) \} , \notag \\
    \{ v_j(k_x) \} \stackrel{TC_2}{=} \{ -v_j(k_x) \} .
\end{align}

Since the system is $T C_2$-symmetric, we can compute the second Stiefel-Whitney number $w_2$ over the whole Brillouin zone \cite{PhysRevLett.121.106403}. However, it is not an independent topological invariant since it is actually equal to the Fu-Kane-Mele invariant $\nu_{\mathrm{FKM}}$. To prove this fact, note that the second Stiefel-Whitney number $w_2$ is equal to the parity of the totally number of crossing at $v_j=\frac{1}{2}$ which is simply $\sum_j [ v^y_j(k_x=0) + v^y_j(k_x=\pi) ] \mod 2$ \footnote{This second Stiefel-Whitney number $w_2$ is over the whole Brillouin zone, compare to the new $\mathbb{Z}_2$-valued invariant $\nu_{new}$ which is the second Stiefel-Whitney number over the effective half Brillouin zone after a deformation of the Hamiltonian.}. The symmetry constraint $\{ v_j(k_x) \} \stackrel{T}{=} \{ v_j(-k_x) \}$ has been used to show that crossing at $v_j=\frac{1}{2}$ at momentum $k_x$ other than $k_x=0$ and $k_x=\pi$ comes in pairs, i.e. $v^y_j(k_x)=\frac{1}{2}$ implies $v^y_{j'}(-k_x)=\frac{1}{2}$ for some Wannier center index $j'$.

\section{General theory} \label{sec: Theory}
In this section, we present a general theory to compute the new topological invariant $\nu_{new}$ under the assumption that three obstruction topological invariants $\nu_{\Gamma X},\nu_{\Gamma Y},\nu_{\mathrm{FKM}}$ vanish. This assumption can be weakened in our homotopic treatment, however, we will keep this assumption to prove that our topological invariant agrees with the cohomology element in K group.

In the following two subsections, we discuss cases of two occupied bands and four occupied bands. We distinguish these two cases based on the differences in their homotopy classification, although their K theory classification is the same.

\subsection{The case of two occupied bands}
In this case, the Wilson loop matrix belong to the $O(2)$ group due to $T C_2$ composite symmetry; its first homotopy group is $\pi_1(O(2)) \cong \mathbb{Z}$. Hence, the homotopy classification is $\mathbb{Z}$ instead of $\mathbb{Z}_2$ in the K theory classification. The $\mathbb{Z}$-valued topological invariant is the Euler number of the system. The Euler number can be written in terms of Berry curvature in a real gauge \cite{PhysRevLett.121.106403}
\begin{align}
    e[BZ] = \frac{1}{2 \pi} \int_{BZ} F_{12} dk_x dk_y,
\end{align}
where $e$ is the Euler class, $[BZ]$ is the fundamental class of the base manifold $BZ$ in the homology group, and their cap product $e[BZ]$ is an integer called the Euler number. From the viewpoint of the Wilson loop spectrum, the Euler number is equal to a protected non-trivial winding, as discussed in Refs. \cite{PhysRevB.99.045140,PhysRevB.100.195135}. This non-trivial winding number can not be changed under a symmetry protected deformation of the Hamiltonian without closing the energy gap. As discussed in Ref. \cite{PhysRevB.100.195135}, when one embeds the system with two occupied bands into a system having four or more occupied bands, the winding number in the Wilson loop spectrum can be changed. However, there is still a $\mathbb{Z}_2$-valued topological number that can not be changed in the case of four occupied bands, and we will discuss this topic in the next subsection.

Now, we want to associate the $\mathbb{Z}$-valued homotopy invariant $e[BZ]$ with the $\mathbb{Z}_2$-valued K theory invariant $\nu_{new}$ when embedding the system into one with more occupied bands. Since the three obstruction invariants vanish, we can deform the Hamiltonian in a symmetric way such that the deformed Hamiltonian is constant on the boundary of the effective half Brillouin zone. During this deformation, the Euler number $e[BZ]$ does not change. After this deformation, the Hamiltonian can be viewed as two Hamiltonians over the two effective half Brillouin zones which are related by TR-symmetry. The Euler number is then expressed as
\begin{align}
    e[\text{BZ}] = & \frac{1}{2 \pi} \int_{\text{BZ}} F_{12} dk_x dk_y \notag \\
    = & \frac{1}{2 \pi} \int_{\text{BZ}^{(1)}_{\frac{1}{2}}} F_{12} dk_x dk_y + \frac{1}{2 \pi} \int_{\text{BZ}^{(2)}_{\frac{1}{2}}} F_{12} dk_x dk_y \notag \\
    = & e[\text{BZ}^{(1)}_{\frac{1}{2}}] + e[\text{BZ}^{(2)}_{\frac{1}{2}}] ,
\end{align}
where $\text{BZ}^{(1)}_{\frac{1}{2}}$ and $\text{BZ}^{(2)}_{\frac{1}{2}}$ are the two effective half Brillouin zone. Since Hamiltonians over $\text{BZ}^{(1)}_{\frac{1}{2}}$ and $\text{BZ}^{(2)}_{\frac{1}{2}}$ are related by TR-symmetry, one can obtain
\begin{align}
    e[\text{BZ}^{(1)}_{\frac{1}{2}}] = e[\text{BZ}^{(2)}_{\frac{1}{2}}].
\end{align}
The K theory topological invariant $\nu_{new}$ is the second Stiefel-Whitney number over the effective half Brillouin zone $\text{BZ}_{\frac{1}{2}}$ which is a mod 2 version of the Euler number \cite{10.2307/j.ctt1b7x751}, that is 
\begin{align}
    \nu_{new} & = w_2[\text{BZ}^{(1)}_{\frac{1}{2}}] \notag \\
    & = e[\text{BZ}^{(1)}_{\frac{1}{2}}] \mod 2 \notag \\
    & = \frac{e[\text{BZ}]}{2} \mod 2.
\end{align}

It should be noted that $e[\text{BZ}^{(1)}_{\frac{1}{2}}]$ is always an integer; in other words, the Euler number $e[\text{BZ}]$ is always an even number. We have made the assumption that $\nu_{\mathrm{FKM}}=0$, which implies that the second Stiefel-Whitney number over the whole Brillouin zone vanishes and the Euler number over the whole Brillouin zone is an even number. Thus $\nu_{new}$ is well-defined.

This `proof' of the correspondence between the $2\mathbb{Z}$-valued Euler number $e[\text{BZ}]$ and the $\mathbb{Z}_2$-valued topological invariant $\nu_{new}$ is not rigorous since the deformation of the Hamiltonian exists only in a stable sense. Due to this $\mathbb{Z}_2$-valued topological invariant $\nu_{new}$ is only useful when the space of two occupied bands is embedded in a space with higher dimensional occupied bands. We will provide rigorous proof of this correspondence in the next subsection, which discusses the case of four occupied bands. 

\subsection{The case of four occupied bands}
\subsubsection{Homotopy invariant} \label{sec: invariant}
We present a new homotopy invariant for the case of four occupied bands in this section. The Wilson loop matrix in this case belong to the $SO(4)$ group due to TR-symmetry and twofold rotation symmetry. In this system, we have a loop of Wilson loop matrices $k_x \mapsto W_{\mathbf{e}_2,\pi}(k_x) : S^1 \rightarrow SO(4)$. The topology information of the system is encoded by this loop since the Wilson loop matrix actually represents the transition function of the system (only the information of the partial polarization along the $x$-direction is lost, and it is encoded in the Wilson loop along another direction). Furthermore, we give a proof in appendix \ref{sec: Wilson} that a homotopy between two Wilson loops induces a homotopy between two Hamiltonians with respect to all symmetries. To study the homotopy between two Wilson loops, we should lift loops in $SO(4)$ to loops in the universal covering group of $SO(4)$, i.e. the $Spin(4)$ group.

There is a two-to-one covering map from the $Sp(1) \times Sp(1)$ group to the $SO(4)$ group, where the $Sp(1)$ group is the one dimensional symplectic group, i.e. the set of quaternion numbers of modulus one. This covering map is 
\begin{align} \label{eq: covering_map}
    p : Sp(1) \times Sp(1) & \longrightarrow  SO(4) \notag \\
    (g,h) & \longmapsto (x \mapsto g^{-1}  x  h),
\end{align}
where $x$ in the expression is a quaternion number, and $x \mapsto g^{-1}  x  h$ is a $\mathbb{R}$-linear map from $\mathbb{R}^4$ to $\mathbb{R}^4$. It is a linear map preserving the norm and having determinant one, hence it belongs to $SO(4)$. Furthermore, this map is a group homomorphism
\begin{align}
    & p((g_1,h_1)  (g_2,h_2)) \notag \\
    = & p((g_1  g_2, h_1  h_2)) \notag \\
    = & x \mapsto g_2^{-1}  g_1^{-1}  x  h_1  h_2 \notag \\
    = & p((g_2,h_2)) \circ p((g_1,h_1)) .
\end{align}

The following theorem \cite{hatcher2002algebraic,bredon1993topology} will be needed in our further discussion.
 \begin{theorem} \label{thm: homotopy}
    Take two maps $f_0,f_1: [0, 2\pi] \rightarrow SO(4)$, such that they are homotoptic, i.e. $f_0 \simeq f_1 \; \mathrm{rel} \; \partial [0,2\pi]$. Let $\tilde{f}_0,\tilde{f}_1: [0, 2\pi] \rightarrow Sp(1) \times Sp(1)$ be liftings of $f_0$ and $f_1$ such that $\tilde{f}_0(0)=\tilde{f}_1(0)$. Then $\tilde{f}_0(2 \pi)=\tilde{f}_1(2 \pi)$ and $\tilde{f}_0 \simeq \tilde{f}_1 \; \mathrm{rel} \; \partial  [0,2\pi]$.
\end{theorem}
The $\mathrm{rel} \; \partial  [0,2\pi]$ in the homotopy notation used in the above theorem means the start and the end point of the path should be fixed, which in our case we simply set as $f_0(0)=f_1(0)=f_0(2 \pi)=f_1(2 \pi)$. This theorem reduces the problem of finding a homotopy between two Wilson loops to the problem of finding a homotopy between liftings of the two Wilson loops.

We first make the assumption that $\nu_{\mathrm{FKM}} = w_2[BZ] = 0$. This assumption implies $[k \mapsto W_{\mathbf{e}_2,\pi}(k)] = 0 \in \pi_1(SO(4))$, which implies that the lifting of the Wilson loop is a loop. That is $\tilde{W}_{\mathbf{e}_2,\pi}(0) = \tilde{W}_{\mathbf{e}_2,\pi}(2 \pi) \in Sp(1) \times Sp(1)$. Otherwise, if $\nu_{\mathrm{FKM}}=1$, $\tilde{W}_{\mathbf{e}_2,\pi}(2 \pi) = -\tilde{W}_{\mathbf{e}_2,\pi}(0)$. By theorem \ref{thm: homotopy}, Wilson loops of these two cases are not homotopic since end points of their liftings are different. This is an obstruction shown by the Fu-Kane-Mele invariant.

Up to now, we have only considered the constraint imposed by $T C_2$ symmetry, which requires the Wilson loop to be real. The constraint imposed by TR-symmetry should be considered as well. The TR-symmetry imposes that
\begin{align} 
    w W_{\mathbf{e}_2,\pi}(k_x) w^{-1} = W^{-1}_{\mathbf{e}_2,\pi}(-k_x),
\end{align}
where $w$ is the sewing matrix of the TR-symmetry, which can be taken to be a constant matrix in general, see Eq. (\ref{eq: W_Tsymm}) in the next section. The homotopy between two Wilson loops should respect this constraint as well. Denote it by $W(k,t)$, where $t$ is a deformation parameter. It should satisfy
\begin{align} \label{eq: wil_constraint}
    w W(k,t) w^{-1} = W^{-1}(-k,t).
\end{align}

Changing the basis if necessary, the sewing matrix of TR-symmetry can be assumed to be $w = \begin{pmatrix}
    0 & -I_{2 \times 2} \\
    I_{2 \times 2} & 0
\end{pmatrix} \in SO(4)$. It maps a quaternion number $x_0+x_1 i + x_2 j + x_3 k$ to $-x_2 -x_3 i +x_0 j + x_1 k$. Since $-x_2 -x_3 i +x_0 j + x_1 k = 1^{-1} \cdot (x_0+x_1 i + x_2 j + x_3 k) \cdot j$, the lifting $\tilde{w}$ of $w$ is $\pm (1,j) \in Sp(1) \times Sp(1)$. The sign of $\tilde{w}$ does not affect the lifting of Eq. (\ref{eq: wil_constraint}), hence we simply take $\tilde{w} = (1,j)$.

Now, we lift this constraint to the covering group $Sp(1) \times Sp(1)$. The lifting requires fixing the start point, and the lift of the left hand side(LHS) of Eq. (\ref{eq: wil_constraint}) should be equal to the lift of the right hand side(RHS) of this equation. The start point of the lift of the LHS is $p^{-1}(w W(0,t) w^{-1}) = \pm \tilde{w}^{-1} \cdot \tilde{W}(0,t) \cdot \tilde{w}$, and we simply choose it to be $\tilde{w}^{-1} \cdot \tilde{W}(0,t) \cdot \tilde{w}$. The start point of the lift of the RHS is $\widetilde{W^{-1}}(0,t)=\pm \tilde{W}(0,t)^{-1}$(we can not choose the sign of RHS since we have already made the sign choice of LHS). Assume $\tilde{W}(0,t)=(x_0+x_1 i+x_2 j+x_3 k,x'_0+x'_1 i +x'_2 j+x'_3 k)$. The start point of LHS is equal to the start point of RHS,
\begin{align}
    & \tilde{w}^{-1} \cdot \tilde{W}(0,t) \cdot \tilde{w} = \pm \tilde{W}(0,t)^{-1} \notag \\
    \Longleftrightarrow & \tilde{W}(0,t) \cdot \tilde{w} = \pm \tilde{w} \cdot \tilde{W}(0,t)^{-1} \notag \\
    \Longleftrightarrow & (x_0+x_1 i+x_2 j+x_3 k,x'_0+x'_1 i +x'_2 j+x'_3 k) \cdot (1,j) \notag \\
    = & \pm (x_0-x_1 i-x_2 j-x_3 k,x'_0+x'_1 i -x'_2 j+x'_3 k) \notag \\
    & \cdot (1,j) . \notag
\end{align}
First let the above equality hold for the `+' sign. This requires $x_1=x_2=x_3=x'_2=0$, which means $\tilde{W}(0,t)=(x_0+x_1 i+x_2 j+x_3 k,x'_0+x'_1 i +x'_2 j+x'_3 k) \in X_0 \bigsqcup Y_0$, where $X_0 = \{ (1,x'_0+x'_1 i + x'_3 k) | (x'_0)^2 + (x'_1)^2 + (x'_3)^2 = 1\}$ and $Y_0 =-X_0= \{ (-1,x'_0+x'_1 i + x'_3 k) | (x'_0)^2 + (x'_1)^2 + (x'_3)^2 = 1\}$. Let $\tilde{W}(0,t) = (1,a+b \, i +d \, k)$, where $a^2+b^2+d^2=1$. Then the matrix form of the Wilson loop $W(0,t)$ which maps $(x_0+x_1 i+x_2 j+x_3 k)$ to $1^{-1} (x_0+x_1 i+x_2 j+x_3 k) (a+b \, i +d \, k)$ is 
\begin{align}
    W(0,t) =
    \begin{pmatrix}
        a & -b & 0 & -d \\
        b & a & d  & 0 \\
        0 & -d & a & b \\
        d & 0 & -b & a
    \end{pmatrix}. \notag
\end{align}
It has eigenvalues $a+\sqrt{1-a^2}\, i$,$a+\sqrt{1-a^2}\, i$,$a-\sqrt{1-a^2}\, i$ and $a-\sqrt{1-a^2}\, i$. Hence, the partial polarization $\nu_{\mathbf{\Gamma Y}}=\sum_j v^y_j(k_x=0) \mod 2 = 0 \mod 2$. Now, let the above equality holds for the `-' sign. This requires $x'_0=x'_1=x'_3=x_0=0$, which means $\tilde{W}(0,t)=(x_0+x_1 i+x_2 j+x_3 k,x'_0+x'_1 i +x'_2 j+x'_3 k) \in X_1 \bigsqcup Y_1$, where $X_1 = \{ (x_1 i + x_2 j + x_3 k,j) | (x_1)^2 + (x_2)^2 + (x_3)^2 = 1 \}$ and $Y_1 = \{ (x_1 i + x_2 j + x_3 k,-j) | (x_1)^2 + (x_2)^2 + (x_3)^2 = 1\}$. Let $\tilde{W}(0,t) = (b \, i + c \, j+d \, k,j)$, where $b^2+c^2+d^2=1$. Then the matrix form of the Wilson loop $W(0,t)$ which maps $(x_0+x_1 i+x_2 j+x_3 k)$ to $(b \, i + c \, j+d \, k)^{-1} (x_0+x_1 i+x_2 j+x_3 k) j$ is 
\begin{align}
    W(0,t) =
    \left(
    \begin{array}{cccc}
    c & d & 0 & -b \\
    d & -c & b & 0 \\
    0 & b & c & d \\
    -b & 0 & d & -c \\
    \end{array}
    \right). \notag
\end{align}
It has eigenvalues $1$,$1$,$-1$ and $-1$. Hence, the partial polarization $\nu_{\mathbf{\Gamma Y}}=\sum_j v^y_j(k_x=0) \mod 2 = 1 \mod 2$. From above, we conclude
\begin{align}
    \tilde{w}^{-1} \cdot \tilde{W}(0,t) \cdot \tilde{w} = \widetilde{W^{-1}}(0,t) = (-1)^{\nu_{\mathbf{\Gamma Y}}} \tilde{W}^{-1}(0,t) .
\end{align}

We have $\widetilde{W^{-1}}(k,t) \tilde{W}(k,t) = p^{-1}(I_{4 \times 4}) = \pm (1,1)$, since $p(\widetilde{W^{-1}}(k,t) \tilde{W}(k,t)) =  W(k,t) W^{-1}(k,t) = I_{4 \times 4}$. The value of $\widetilde{W^{-1}}(k,t) \tilde{W}(k,t)$ continuously depends on $k$, which implies $\widetilde{W^{-1}}(k,t) \tilde{W}(k,t) = \widetilde{W^{-1}}(0,t) \tilde{W}(0,t)$. In the $\nu_{\mathbf{\Gamma Y}} = 0 \mod 2$ case, we have $\widetilde{W^{-1}}(0,t) \tilde{W}(0,t) = (1,1)$, while in the $\nu_{\mathbf{\Gamma Y}} = 1 \mod 2$ case, we have $\widetilde{W^{-1}}(0,t) \tilde{W}(0,t) = (-1,-1)$. Combining these relations, from Eq. (\ref{eq: wil_constraint}) one obatins
\begin{align} \label{eq: lift_constraint}
    & p^{-1}(w W(k,t) w^{-1})  = p^{-1}(W^{-1}(-k,t)) \notag \\
    \Leftrightarrow & 
    \tilde{w}^{-1} \cdot \tilde{W}(k,t) \cdot \tilde{w} = (-1)^{\nu_{\mathbf{\Gamma Y}}} \overline{\tilde{W}(-k,t)}, 
\end{align}
where the bar on the right side of the equality means the conjugation of quaternion numbers, since for a modulus 1 quaternion number $a$ we have $a^{-1}=\bar{a}$; meanwhile, $\widetilde{W^{-1}}(k) \tilde{W}(k) = \pm (1,1)$ is used for different cases. We emphasize here that the vanishing of the Fu-Kane-Mele invariant $\nu_{\mathbf{FKM}}$ implies $\sum_j v^y_j(k_x=0) \mod 2 = \sum_j v^y_j(k_x=\pi) \mod 2$, which implies that Eq. (\ref{eq: lift_constraint}) has the same sign at two time-reversal invariant points $k_x=0$ and $k_x=\pi$, i.e. $(-1)^{\nu_{\mathbf{\Gamma Y}}}=(-1)^{\sum_j v^y_j(k_x=0)} = (-1)^{\sum_j v^y_j(k_x=\pi)}$. We will discuss the non-zero Fu-Kane-Mele invariant case for comparison at the end of this section.

We denote the time-reversal operator acting on $Sp(1) \times Sp(1)$ by $T$,
\begin{align}
    T : Sp(1) \times Sp(1) & \longrightarrow Sp(1) \times Sp(1) \notag \\
    (u,v) & \longmapsto \overline{\tilde{w}^{-1} \cdot (u,v) \cdot \tilde{w}}.
\end{align}
Then Eq. (\ref{eq: lift_constraint}) can be written in terms of $T$ as
\begin{align}
    T(\tilde{W}(k,t)) = (-1)^{\nu_{\mathbf{\Gamma Y}}} \tilde{W}(-k,t).
\end{align}
Note that for $k=0$ or $k=\pi$, $\tilde{W}(k,t)$ should be fixed(or anti-fixed for the non-zero partial polarization case) by the time-reversal operator $T$ for any $t \in [0,1]$, i.e. $T(\tilde{W}(0,t)) = (-1)^{\nu_{\mathbf{\Gamma Y}}} \tilde{W}(0,t)$ and $T(\tilde{W}(\pi,t)) = (-1)^{\nu_{\mathbf{\Gamma Y}}} \tilde{W}(\pi,t)$. Hence, we should study the fixed(anti-fixed) point set of the time-reversal operator $T$.

If $\nu_{\mathbf{\Gamma Y}} = 0 \mod 2$, Eq. (\ref{eq: lift_constraint}) becomes $T(u,v)=(u,v)$ and the solution of it is $x_1=x_2=x_3=x'_2=0$. Hence, the fixed point set of the map $T$ in $Sp(1) \times Sp(1)$ is 
\begin{align}
    & (Sp(1) \times Sp(1))^T \notag \\
    = & \{ (1,x'_0+x'_1 i + x'_3 k) | (x'_0)^2 + (x'_1)^2 + (x'_3)^2 = 1 \} \notag \\
    & \bigsqcup \{ (-1,x'_0+x'_1 i + x'_3 k) | (x'_0)^2 + (x'_1)^2 + (x'_3)^2 = 1\} \notag \\
    = & X_0 \bigsqcup Y_0,
\end{align}
where $X_0 = \{ (1,x'_0+x'_1 i + x'_3 k) | (x'_0)^2 + (x'_1)^2 + (x'_3)^2 = 1\}$ and $Y_0 =-X_0= \{ (-1,x'_0+x'_1 i + x'_3 k) | (x'_0)^2 + (x'_1)^2 + (x'_3)^2 = 1\}$ as before. Note that $X_0$ and $Y_0$ are disconnected in $Sp(1) \times Sp(1)$, and each of $X_0,Y_0$ is a two dimensional sphere. 

The lifted Wilson loop has a global sign ambiguity, i.e. $\pm \tilde{W}(k)$ are both lifts of $W(k)$. We always set the start point of the lifted Wilson loop to be in $X_0$ to avoid this sign ambiguity. The midpoint of the lifted Wilson loop can be either in $X_0$ or in $Y_0$. The simplest trivial phase has a constant Wilson loop $k \mapsto I_{4 \times 4}$. The lifting of this Wilson loop is also a constant loop which is $k \mapsto (1,1)$. At $k = \pi$, it belong to the fixed point subset $X_0$. Hence, we can conclude the following theorem: 
\begin{theorem} \label{thm: trivial}
    The TR-symmetric and twofold rotation symmetric Hamiltonian with four occupied bands with vanishing $\nu_{\mathrm{FKM}}$ and $\nu_{\mathbf{\Gamma Y}}$ is topologically trivial if the lifting $\tilde{W}_{\mathbf{e}_2,\pi}(k_x)$ of its Wilson loop $W_{\mathbf{e}_2,\pi}(k_x)$ satisfies $\tilde{W}_{\mathbf{e}_2,\pi}(k_x=\pi) \in X_0$. If $\tilde{W}_{\mathbf{e}_2,\pi}(k_x=\pi) \in Y_0$, it is non-trivial.
\end{theorem}
We present a proof of this theorem in appendix \ref{sec: proof}. The basic idea behind this is that the disconnectedness of the fixed point set is an obstruction to deform the Wilson loop to a trivial loop. It implies that even in a case of four occupied bands, the spectrum of some non-trivial Wilson loop can not unwind, although its winding number can change. Hence, the first component in tuple $\tilde{W}_{\mathbf{e}_2,\pi}(k_x=\pi)$ is a $\mathbb{Z}_2$-valued homotopy invariant distinguishing the trivial phase and the non-trivial phase.

If $\nu_{\mathbf{\Gamma Y}} = 1 \mod 2$, Eq. (\ref{eq: lift_constraint}) becomes $T(u,v)=-(u,v)$ and the solution of it is $x'_0=x'_1=x'_3=x_0=0$. Hence, the anti-fixed point set of the map $T$ in $Sp(1) \times Sp(1)$ is 
\begin{align}
    & (Sp(1) \times Sp(1))^{-T} \notag \\
    = & \{ (x_1 i + x_2 j + x_3 k,j) | (x_1)^2 + (x_2)^2 + (x_3)^2 = 1 \} \notag \\
    & \bigsqcup \{ (x_1 i + x_2 j + x_3 k,-j) | (x_1)^2 + (x_2)^2 + (x_3)^2 = 1\} \notag \\
    = & X_1 \bigsqcup Y_1,
\end{align}
where $X_1 = \{ (x_1 i + x_2 j + x_3 k,j) | (x_1)^2 + (x_2)^2 + (x_3)^2 = 1 \}$ and $Y_1 = \{ (x_1 i + x_2 j + x_3 k,-j) | (x_1)^2 + (x_2)^2 + (x_3)^2 = 1\}$ as before. Note that $X_1$ and $Y_1$ are disconnected in $Sp(1) \times Sp(1)$, and each of $X_1,Y_1$ is a two dimensional sphere.

In this case we always fix the start point of the lifted Wilson loop to be in $X_1$ to avoid the sign ambiguity. The simplest trivial phase in this case has a constant Wilson loop $k \mapsto \mathbf{diag}(1,-1,1,-1)$, whose lift is also a constant loop, which is $k \mapsto (j,j)$. At $k = \pi$, it belong to the fixed point subset $X_1$. Thus we can conclude the following theorem:
\begin{theorem} 
    The TR-symmetric and twofold rotation symmetric Hamiltonian with four occupied bands with vanishing $\nu_{\mathrm{FKM}}$ and non-zero $\nu_{\mathbf{\Gamma Y}}$ is topologically trivial if the lifting $\tilde{W}_{\mathbf{e}_2,\pi}(k_x)$ of its Wilson loop $W_{\mathbf{e}_2,\pi}(k_x)$ satisfies $\tilde{W}_{\mathbf{e}_2,\pi}(k_x=\pi) \in X_1$. If $\tilde{W}_{\mathbf{e}_2,\pi}(k_x=\pi) \in Y_1$, it is non-trivial.
\end{theorem}

The proof of this theorem is similar to the proof of theorem \ref{thm: trivial}.

Now, we consider the case in which the Fu-Kane-Mele invariant $\nu_{\mathbf{FKM}}$ is non-zero. In this case, we lift the Wilson loop starting at $k=0$ along two directions, i.e., we lift along the $(-k)$-direction to obtain a path whose parameter $k$ is in $[-\pi,0]$, and we lift along the $(+k)$-direction to obtain a path whose parameter $k$ is in $[0,\pi]$. As a result, we obatin a lifted Wilson loop $\tilde{W}(k) \in Sp(1) \times Sp(1)$, where $k \in [-\pi,\pi]$. Since the Fu-Kane-Mele invariant $\nu_{\mathbf{FKM}}$ is non-zero, the lifted Wilson loop satisfies $\tilde{W}(\pi) = -\tilde{W}(-\pi)$. Hence, the lifted Wilson loop at $k=\pi$ is related to the Wilson loop at $k=-\pi$ via the relation (\ref{eq: lift_constraint}) instead of to itself. If $\nu_{\mathbf{\Gamma Y}}=\sum_j v^y_j(k=0) = 0 \mod 2$, then $\sum_j v^y_j(k=\pi) = 1 \mod 2$. Further, this implies $\tilde{W}(k=\pi) \in X_1 \bigsqcup  Y_1$ by the same argument we made at $k=0$. Meanwhile, we have $\tilde{W}(k=0) \in X_0 \bigsqcup  Y_0$. In the following, we explain why the lifted Wilson loops at $k=0$ and $k=\pi$ belong to different sets. Because our sign convention is made at the start point $k=0$ and the sign in the relation (\ref{eq: lift_constraint}) which depends on $\sum_j v^y_j(k=0) = 0 \mod 2$ is `$+$', the relation (\ref{eq: lift_constraint}) becomes
\begin{align}
    \tilde{w}^{-1} \cdot \tilde{W}(k) \cdot \tilde{w} = \overline{\tilde{W}(-k)} .
\end{align}
We consider the lifted Wilson loop at $k=0$ and $k=\pi$. Using the above relation we obtain
\begin{align}
    \tilde{w}^{-1} \cdot \tilde{W}(0) \cdot \tilde{w} & = \overline{\tilde{W}(0)}, \notag \\
    \tilde{w}^{-1} \cdot \tilde{W}(\pi) \cdot \tilde{w} & = \overline{\tilde{W}(-\pi)}. \notag
\end{align}
Due to the assumption that the Fu-Kane-Mele invariant $\nu_{\mathbf{FKM}}$ is non-zero, the relation $\tilde{W}(\pi) = -\tilde{W}(-\pi)$ holds. Hence, following relations hold
\begin{align}
    \tilde{w}^{-1} \cdot \tilde{W}(0) \cdot \tilde{w} & = \overline{\tilde{W}(0)}, \notag \\
    \tilde{w}^{-1} \cdot \tilde{W}(\pi) \cdot \tilde{w} & = -\overline{\tilde{W}(\pi)}, \notag
\end{align}
which imply $\tilde{W}(k=0) \in X_0 \bigsqcup  Y_0$ and $\tilde{W}(k=\pi) \in X_1 \bigsqcup  Y_1$.

To avoid sign ambiguity, if $\sum_j v^y_j(k=0) = 0 \mod 2$, we fix the start point of the lifted Wilson loop to be in $X_0$ and otherwise we fix the start point of the lifted Wilson loop to be in $X_1$. By the above discussion, we immediately conclude the following theorem:
\begin{theorem} 
    Consider the TR-symmetric and twofold rotation symmetric Hamiltonian with four occupied bands with non-vanishing $\nu_{\mathrm{FKM}}$. If $\sum_j v^y_j(k=\pi) = 0 \mod 2$, phases with different $\nu_{new}$ invariants are distinguished by two cases: $\tilde{W}(\pi) \in X_0$ and $\tilde{W}(\pi) \in Y_0$. Otherwise, if it is in the case $\sum_j v^y_j(k=\pi) = 1 \mod 2$, phases with different $\nu_{new}$ invariants are distinguished by two cases: $\tilde{W}(\pi) \in X_1$ and $\tilde{W}(\pi) \in Y_1$.
\end{theorem}

\subsubsection{Examples} \label{sec: wil_examples}
In the following, we will prove that the above homotopy classification agrees with the K theory classification. We show this using two typical examples with vaninshing $\nu_{new}$,$\nu_{\mathbf{\Gamma X}}$ and $\nu_{\mathbf{\Gamma Y}}$. Per the discussion in appendix \ref{sec: can_form}, all examples with vanishing $\nu_{new}$, $\nu_{\mathbf{\Gamma X}}$ and $\nu_{\mathbf{\Gamma Y}}$ can be deformed to these two typical examples. Thus we deform each system with vanishing $\nu_{new}$, $\nu_{\mathbf{\Gamma X}}$ and $\nu_{\mathbf{\Gamma Y}}$ into these two examples, and we check that the K theory invariant in these two examples agrees with our homotopy invariant. Since the K theory invariant and our homotopy invariant are both homotopically invariant, i.e., they are not changed under deformation, these proves our homotopy classification agrees with the K theory classification. 

We first distinguish these two examples using two symmetry categories. Each system in the first category has two TR-symmetry related $TC_2$-symmetric channels. It can be expressed as
\begin{align} \label{eq: category1}
    E & = E_{\mathrm{I}} \oplus E_{\mathrm{II}} \notag \\
    TC_2 E_{\mathrm{I,II}} & \subseteq E_{\mathrm{I,II}} \notag \\
    T E_{\mathrm{I}} & \subseteq E_{\mathrm{II}} \notag \\
    T E_{\mathrm{II}} & \subseteq E_{\mathrm{I}},
\end{align}
where $E$ is the space of occupied bands, and $E_{\mathrm{I}}$ and $E_{\mathrm{II}}$ are two channels.
But each system in the second category is characterized by 
\begin{align} \label{eq: category2}
    E & = E_{\mathrm{I}} \oplus E_{\mathrm{II}} \notag \\
    TC_2 E_{\mathrm{I,II}} & \subseteq E_{\mathrm{I,II}} \notag \\
    T E_{\mathrm{I}} & \subseteq E_{\mathrm{I}} \notag \\
    T E_{\mathrm{II}} & \subseteq E_{\mathrm{II}}.
\end{align}
Note that each system in these two categories has TR-related channels. In appendix \ref{sec: Decomposition}, we present a numerical method based on the Wilson loop for TR-related channels decomposition. In the following we will show examples belonging to these two categories that may have the same Wilson loop spectrum, but their topological classes(trivial or non-trivial) is different.

We first consider the following example. The Wilson loop is 
\begin{align}
    & W_{\mathbf{e}_2,k_y=\pi}(k_x) \notag \\
    = & 
    \begin{pmatrix}
        \cos n k_x & \sin n k_x & 0 & 0 \\
        -\sin n k_x & \cos n k_x & 0 & 0 \\
        0 & 0 & \cos n k_x & \sin n k_x \\
        0 & 0 & -\sin n k_x & \cos n k_x
    \end{pmatrix},
\end{align}
where $n$ is an integer,
and the sewing matrix of the TR-symmetry is
\begin{align}
    w=\begin{pmatrix}
        0 & -I_{2 \times 2} \\
        I_{2 \times 2} & 0
    \end{pmatrix}.
\end{align}
Since the Wilson loop is block diagonal and the sewing matrix $w$ is block off-diagonal, the Wannier bands can be decomposed to two channels such that they are TR-related and each is $TC_2$-symmetric. 

We calculate the lifting of the Wilson loop $\tilde{W}_{\mathbf{e}_2,k_y=\pi}(k_x) = (\cos n k_x + i \sin n k_x, 1)$. Hence the midpoint $\tilde{W}_{\mathbf{e}_2,k_y=\pi}(k_x=\pi)$ is equal to $((-1)^n, 1)$, which belongs to $X_0$ for even $n$ and to $Y_0$ for odd $n$. Then by theorem \ref{thm: trivial}, for even $n$ the system is topologically trivial and for odd $n$ it is topologically non-trivial. In other words, the new homotopy invariant is expressed by $n \mod 2$.

\begin{figure} 
    \centering
    \subfigure[]{
        \centering
        \includegraphics[width=.4\columnwidth]{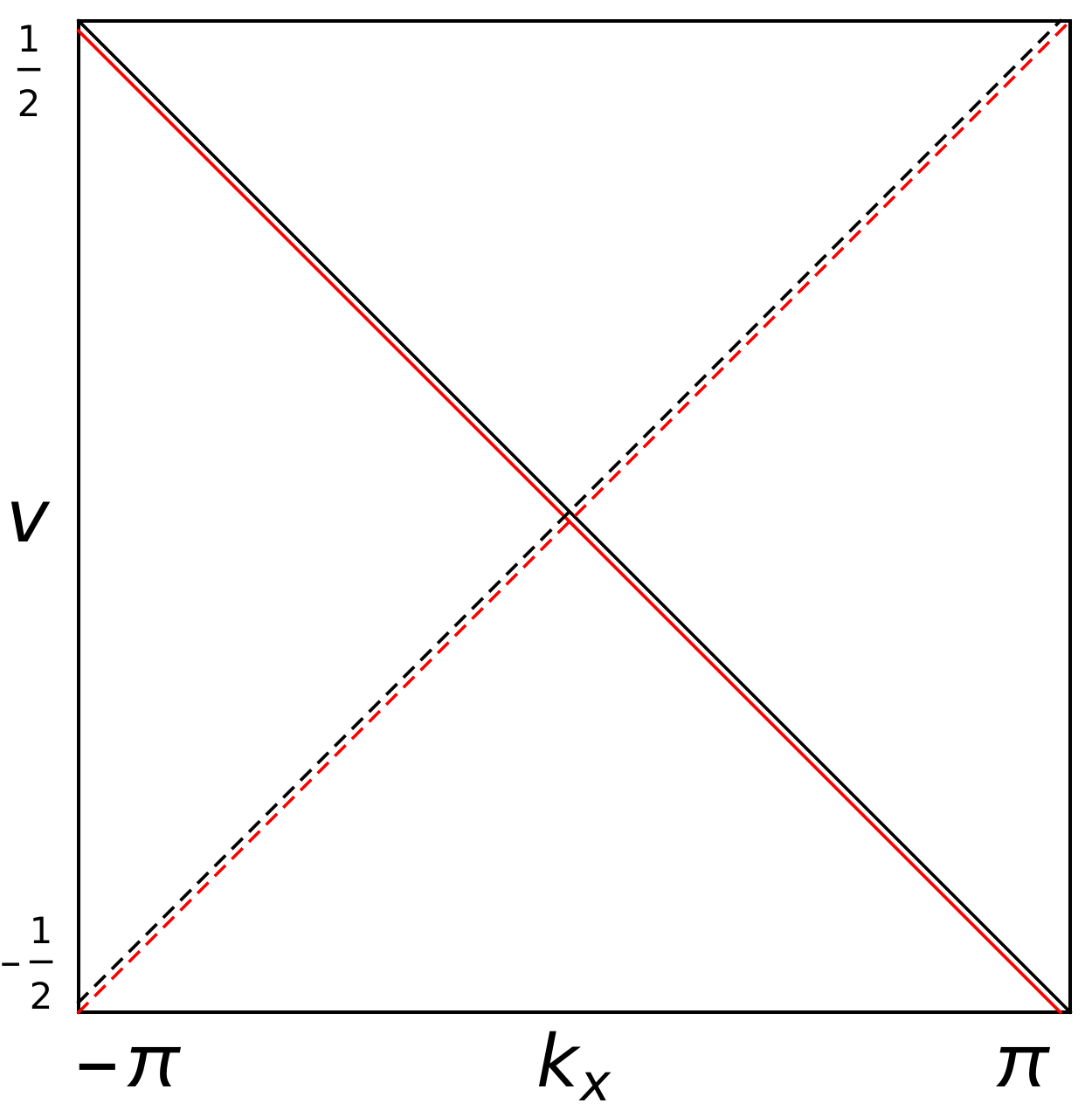}
    }
    \subfigure[]{
        \centering
        \includegraphics[width=.4\columnwidth]{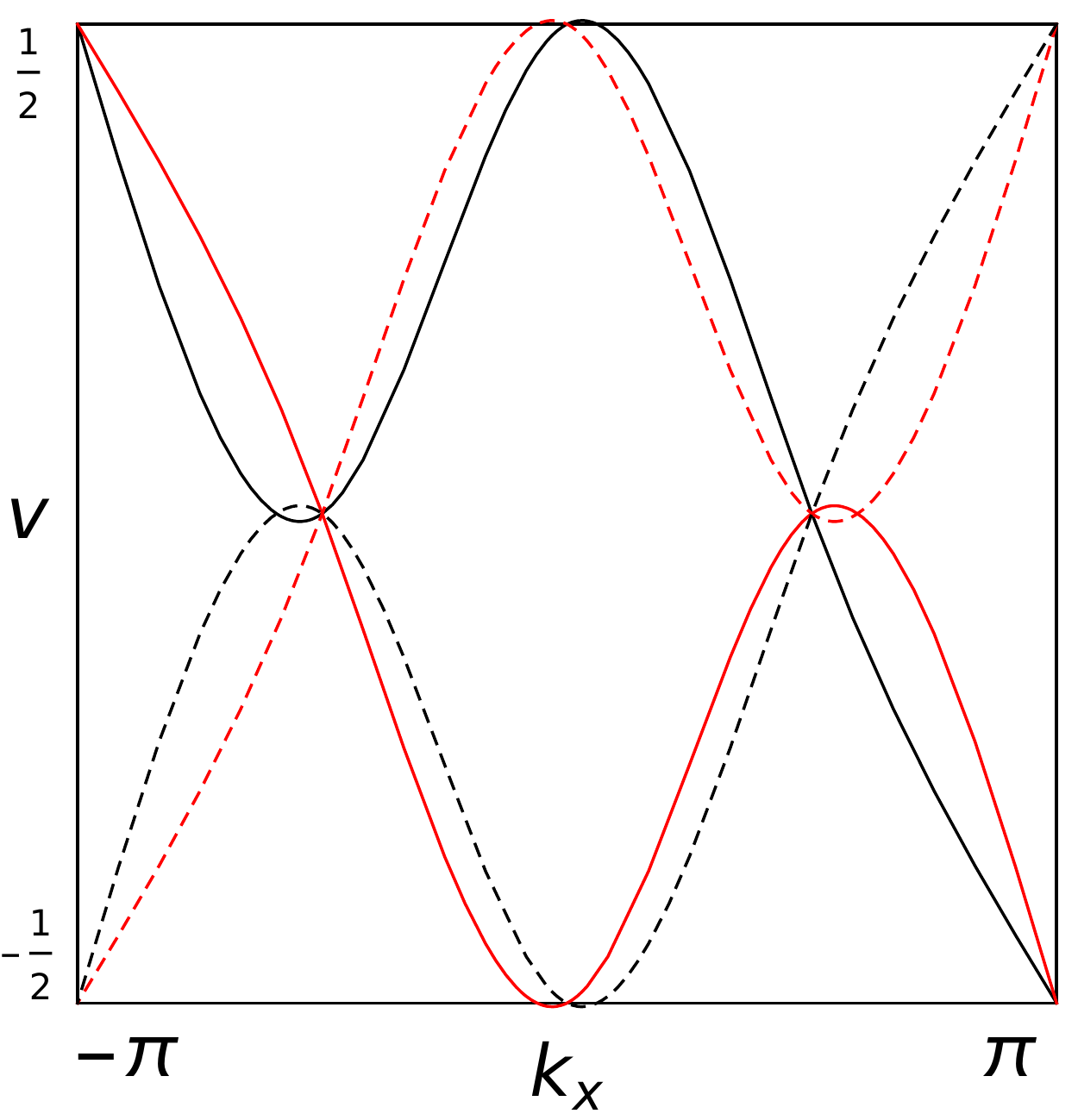}
    }
    \caption[]{(a) The original Wilson loop spectrum. There are four Wannier bands label by black, red, dashed black and dashed red. The black band coincides with the red band and the black dashed band coincides with the red dashed band. (b) The deformed Wilson loop spectrum. The winding number of each Wannier band is not changed. Furthermore, symmetries of Wannier bands are preserved.} \label{fig: deform}
\end{figure}

Without loss of generality, we prove the equivalence of this homotopy topological invariant and the K theory topological invariant $\nu_{new}$ for the $n=1$ case. The Wilson loop spectrum is shown in Figure \ref{fig: deform}(a). It can be seen that the spectrum is not periodic over the effective half Brillouin zone while the only known method to calculate the K theory invariant is to deform the system to a system whose Wilson loop spectrum is periodic over the effective half Brillouin zone and preserving all symmetries. The K theory invariant is not changed during the deformation (it is homotopically invariant), and after the deformation it is the second Stiefel-Whitney number over the effective half Brillouin zone. Such a deformation exists, and the Hamiltonian can be deformed to a Hamiltonian that is constant on the boundary of the effective half Brillouin zone since all three obstruction invariants are zero. We present a deformation in the Wilson loop spectrum level such that the winding number of each Wannier band is not changed. We present the deformed Wilson spectrum in Figure \ref{fig: deform}(b). The deformation preserves all symmetries of the system. Note that if we denote the space with black Wannier bands as channel I, and the space with red Wannier bands as channel II, then these two channels are TR-related and each channel is $TC_2$-symmetric.
Note that on the left effective half Brillouin zone($k_x \in [-\pi,0]$), the black Wannier bands have winding number 0 and the red Wannier bands have winding number 1. However, the roles of the black Wannier bands and the red Wannier bands are exchanged on the right effective half Brillouin zone, i.e., the black Wannier bands have winding number 1 and the red Wannier bands have winding number 0. Hence, the total second Stiefel-Whitney number $\nu_{new}=w_2[BZ_{\frac{1}{2}}]$ over the effective half Brillouin zone is 1, which implies that this phase is non-trivial. Thus we conclude that the homotopy invariant we obtained above agrees with the K theory invariant $\nu_{new}$ in this example and in all systems.

Now, we elaborate on another example which behaves differently from the first example. The Wilson loop is 
\begin{align} \label{eq: wil_example2}
    & W_{\mathbf{e}_2,k_y=\pi}(k_x) \notag \\
    = & 
    \begin{pmatrix}
        \cos m k_x & \sin m k_x & 0 & 0 \\
        -\sin m k_x & \cos m k_x & 0 & 0 \\
        0 & 0 & \cos n k_x & \sin n k_x \\
        0 & 0 & -\sin n k_x & \cos n k_x
    \end{pmatrix},
\end{align}
where $m$ and $n$ are integers, and $m+n$ is even due to the requirement of vanishing of the Fu-Kane-Mele invariant. The sewing matrix of the TR-symmetry is 
\begin{align}
    w = \begin{pmatrix}
        0 & -1 & 0 & 0 \\
        1 & 0 & 0 & 0\\
        0 & 0 & 0 & -1\\
        0 & 0 & 1 & 0
    \end{pmatrix}.
\end{align}
Since both $W_{\mathbf{e}_2,k_y=\pi}(k_x)$ and $w$ are block diagonal, the space of occupied bands can be decomposed into two $T,C_2$-symmetric subspaces with two occupied bands, which implies that this example belongs to the second category. We apply a basis transformation on the sewing matrix to the previous standard form $w=\begin{pmatrix}
    0 & -I_{2 \times 2} \\
    I_{2 \times 2} & 0
\end{pmatrix}$. Under this basis transformation, the Wilson loop becomes
\begin{align} \label{eq: wil_example2_0}
    & W_{\mathbf{e}_2,k_y=\pi}(k_x) \notag \\
    = & 
    \begin{pmatrix}
        \cos m k_x & 0 & \sin m k_x & 0 \\
        0 & \cos n k_x & 0 & \sin n k_x \\
        -\sin m k_x & 0 & \cos m k_x & 0 \\
        0 & -\sin n k_x & 0 & \cos n k_x
    \end{pmatrix}.
\end{align}
By solving the lifting equation, the lifting of this Wilson loop is
\begin{align} \label{eq: lifted_wil2}
    & \tilde{W}_{\mathbf{e}_2,k_y=\pi}(k_x) \notag \\
    = & \bigg ( \frac{\cos m k_x + \cos n k_x}{2 \cos(\frac{m+n}{2} k_x)} + j \frac{\sin m k_x - \sin n k_x}{2 \cos(\frac{m+n}{2} k_x)}, \notag \\
    & \cos(\frac{m+n}{2} k_x) - j \sin (\frac{m+n}{2} k_x)  \bigg ),
\end{align}
which can be directly verified by 
\begin{align}
    \left[ \frac{\cos m k_x + \cos n k_x}{2 \cos(\frac{m+n}{2} k_x)} -  j \frac{\sin m k_x - \sin n k_x}{2 \cos(\frac{m+n}{2} k_x)} \right] \notag \\
    \cdot (x_0 + x_1 i + x_2 j + x_3 k) \notag \\
    \cdot [ \cos(\frac{m+n}{2} k_x) - j \sin (\frac{m+n}{2} k_x)] = \notag \\
    (1, i, j, k) \cdot W_{\mathbf{e}_2,k_y=\pi}(k_x) \cdot (x_0, x_1, x_2, x_3)^T . \notag
\end{align}
The midpoint of the lifting is $\tilde{W}_{\mathbf{e}_2,k_y=\pi}(k_x = \pi) =\bigg  (\frac{(-1)^m+(-1)^n}{2 (-1)^{\frac{m+n}{2}}},* \bigg) = \bigg (\frac{(-1)^m[1+(-1)^{n-m}]}{2 (-1)^{\frac{m+n}{2}}},* \bigg) = ((-1)^m(-1)^{\frac{-m-n}{2}},*) = ((-1)^{\frac{m-n}{2}},*) $. Hence the system is in a trivial phase if and only if $\frac{m-n}{2}$ is even. In other words, the new homotopy invariant is expressed as $\frac{m-n}{2} \mod 2$. Note that $\frac{m-n}{2}$ is ensured to be an integer since $m+n$ is even due to the requirement of vanishing of the Fu-Kane-Mele invariant. 

Now, consider the $m=n=1$ case. The Wilson loop spectrum is identical to the first example with $n=1$, but their topological classes are different, i.e., the first example is non-trivial and this example is trivial. This is because these two examples belong to different categories described above, and this example can reduce to the problem of two subsystems with two occupied bands while the first example can not.

Now, we prove the equivalence of this homotopy invariant and the K theory invariant $\nu_{new}$ in this example. Consider a non-trivial phase, say, $m=1,n=3$. It can be seen that the Wilson loop spectrum is not periodic over the effective half Brillouin zone, and we can not deform the Wilson loop spectrum with respect to symmetries without alternating the winding number of each Wannier band. However, we have proved that the homotopy class of this example only depends on $m-n$. Hence, this phase is equivalent to a phase with $m=2+2j,n=4+2j,j \in \mathbb{Z}$. The Wilson loop with $m=2+2j,n=4+2j$ has a periodic spectrum over the effective half Brillouin zone, and the K theory invariant is calculated as $\nu_{new}= w_2[BZ_{\frac{1}{2}}] = \frac{m}{2} + \frac{n}{2} \mod 2 = 1 \mod 2$ which implies this phase is non-trivial. Therefore, the homotopy invariant $pr_1(\tilde{W}_{\mathbf{e}_2,\pi}(k_x = \pi))$ agrees with the K theory invariant $\nu_{new}$, where $pr_1$ in this expression takes the first component of the tuple. From now on, we identify our homotopy invariant with the K theory invariant $\nu_{new}$ and simply denote our homotopy invariant by $\nu_{new}$.

In this example, the system can be decomposed to two subsystems with two occupied bands. Therefore, we can rigorously prove the conclusion $\nu_{new}=\frac{e[BZ]}{2} \mod 2$ for the case of two occupied bands. Denote occupied spaces of two subsystems by $E_1$ and $E_2$. Let $m,n$ be even so that each subsystem has a vanishing Fu-Kane-Mele invariant. Euler numbers of two subsystems are given by $e(E_1)[BZ]=m$ and $e(E_2)[BZ]=n$ respectively. We simply set $n=0$ so that the subsystem $E_2$ is trivial. Due to linearity, the invariant of this four band system is sum of two invariants of subsystems, i.e. $\nu_{new}(E_1 \oplus E_2) = \nu_{new}(E_1) + \nu_{new}(E_2)$. Relations $\nu_{new}(E_1 \oplus E_2) = \frac{m-0}{2} \mod 2$ and $\nu_{new}(E_2)=0 \mod 2$ imply $\nu_{new}(E_1) = \frac{m}{2} = \frac{e(E_1)[BZ]}{2} \mod 2$. Conversely, this result and linearity imply $\nu_{new}(E_1 \oplus E_2) = \nu_{new}(E_1) + \nu_{new}(E_2) = \frac{m}{2} + \frac{n}{2} = \frac{m-n}{2} \mod 2$ for non-zero $n$, where in the last equality the  fact that $\frac{m-n}{2} \mod 2= \frac{m+n}{2} \mod 2$ for even $m,n$ is used.

We summarize the quantitive results of this section in this paragraph. Examples belonging to the first category are characterized by a single Wilson loop winding number $n$, and the new topological invariant is expressed by $\nu_{new}= n \mod 2$. Meanwhile, examples belonging to the second category are characterized by two Wilson loop winding numbers $m$ and $n$, where each winding number is the Euler number of a $T,C_2$-symmetric subsystem, and the new topological invariant is expressed by $\nu_{new}=\frac{m-n}{2} \mod 2$.

\section{An example of Hamiltonian} \label{sec: ham_Example}
\begin{figure}
    \includegraphics[width=.8\columnwidth]{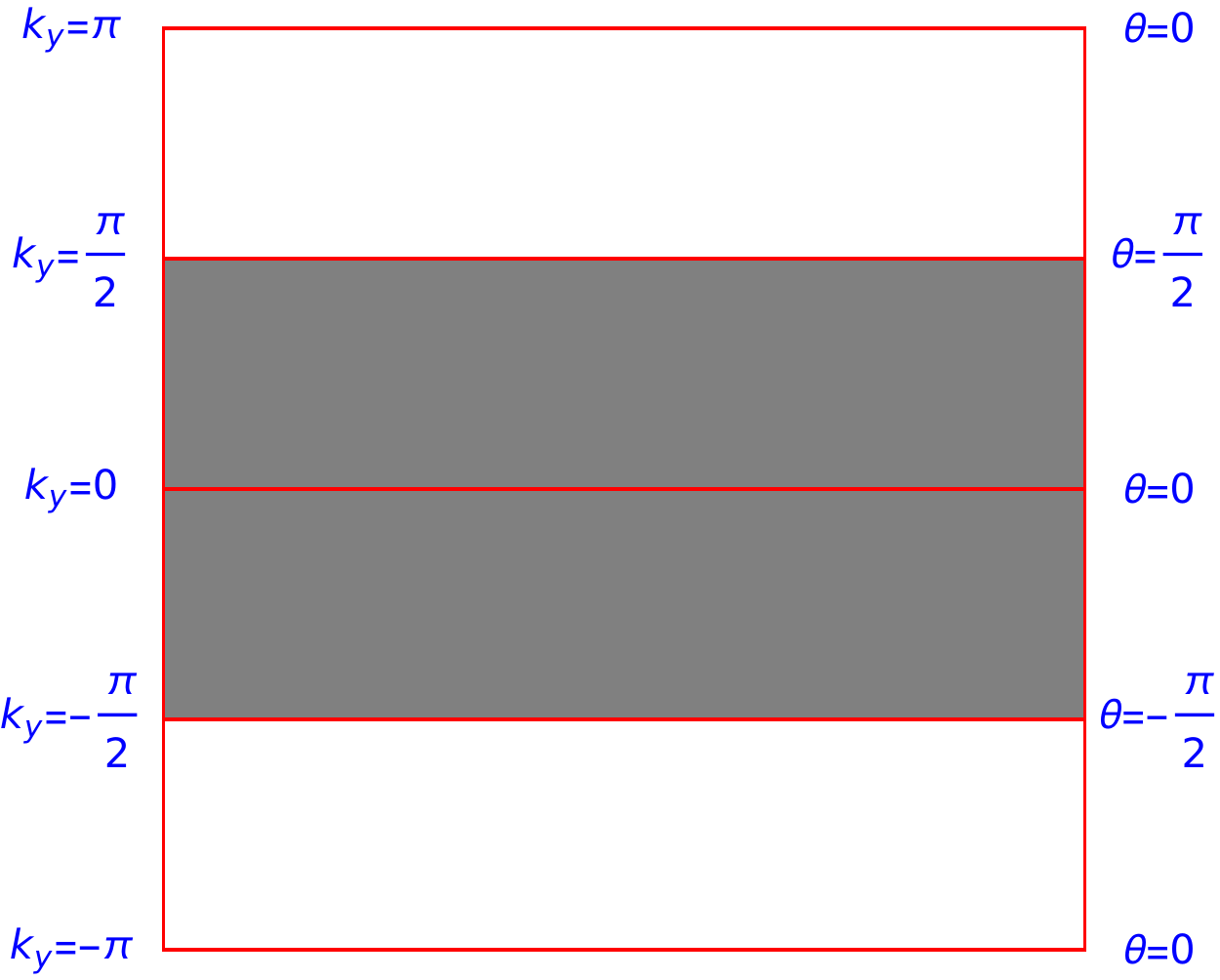}
    \caption[The map from the value of $k_y$ to the value of $\theta$]{The map from the value of $k_y$ to the value of $\theta$. The left side of the square(Brillouin zone) shows the value of $k_y$, and the right side of the square shows the value of $\theta$. The value of $\theta$ between two labeled values of $\theta$ is linear interpolated. The value of $q(k_x)$ in Eq. (\ref{eq:ham}) is equal to $q_1(k_x)$ in gray region, and $q_0(k_x)$ in white region.} \label{fig:thetamap}
\end{figure}

In this section, we give an example of Hamiltonian on momentum space whose Wilson loop operator can be analytically computed. The basic idea of this example is that the Berry connection of this system is flat (zero), hence the only contribution to the Wilson loop operator is the transition function between two patches. The eight bands Hamiltonian is given in the form of the image of a dimension raising isomorphism,
\begin{align} \label{eq:ham}
    H(k_x,k_y) = \cos \theta \begin{pmatrix}
        0 & q(k_x) \\
        q^T(k_x) & 0
    \end{pmatrix} + \sin \theta \begin{pmatrix}
        I_{4 \times 4} & 0\\
        0 & -I_{4 \times 4}
    \end{pmatrix},
\end{align}
where $\theta \in [-\pi/2,\pi/2]$, and $q(k_x)$ in this expression equals $q_1(k_x)$ for $k_y \in [-\frac{\pi}{2},\frac{\pi}{2}]$ and equals $q_0(k_x)$ for $k_y \in [-\pi,-\frac{\pi}{2}] \cup [\frac{\pi}{2},\pi]$. The map from the value of $k_y$ to the value of $\theta$ is shown in Figure \ref{fig:thetamap}.
For $k_y \in [-\frac{\pi}{2},\frac{\pi}{2}]$ (gray region in Figure \ref{fig:thetamap}), 
\begin{align}
    q(k_x)=q_1(k_x) =
    \begin{pmatrix}
        0 & 0 & \cos n k_x & -\sin n k_x \\
        0 & 0 & \sin n k_x & \cos n k_x \\
        \cos n k_x & -\sin n k_x & 0 & 0 \\
        \sin n k_x & \cos n k_x & 0 & 0
    \end{pmatrix},
\end{align}
where $n$ is an integer.
As for $k_y \in [-\pi,-\frac{\pi}{2}] \cup [\frac{\pi}{2},\pi]$ (white region in Figure \ref{fig:thetamap}),
\begin{align}
    q(k_x)=q_0(k_x) =
    \begin{pmatrix}
        0 & 0 & 1 & 0 \\
        0 & 0 & 0 & 1 \\
        1 & 0 & 0 & 0 \\
        0 & 1 & 0 & 0
    \end{pmatrix}.
\end{align}
This Hamiltonian is continuous and periodic on the Brillouin zone.
This system has a TR-symmetry operator $T$ and a twofold rotation symmetry operator $C_2$, which are
\begin{align}
    T = 
    \begin{pmatrix}
        0 & -\sigma_z \otimes \sigma_0\\
        \sigma_z \otimes \sigma_0 & 0
    \end{pmatrix} \mathcal{K}, \notag \\
    C_2 =
    \begin{pmatrix}
        0 & \sigma_z \otimes \sigma_0 \\
        -\sigma_z \otimes \sigma_0 & 0
    \end{pmatrix},
\end{align}
where $\mathcal{K}$ is the complex conjugation operator, and the composite symmetry $T C_2$ is equal to $\mathcal{K}$.

Assume this system is half-filling, which makes the system a gapped system. The Wilson loop operator can be analytically computed in this system, and we give the computation in appendix \ref{sec: Wilson}. The result is expressed in terms of $q_1(k_x)$ and $q_0(k_x)$
\begin{align} \label{eq: wil}
    & W_{\mathbf{e}_2,k_y=\pi}(k_x) \notag \\
    = & q_1^T(k_x) q_0^T(k_x)^{-1} \notag \\
    = & 
    \begin{pmatrix}
        \cos n k_x & \sin n k_x & 0 & 0 \\
        -\sin n k_x & \cos n k_x & 0 & 0 \\
        0 & 0 & \cos n k_x & \sin n k_x \\
        0 & 0 & -\sin n k_x & \cos n k_x
    \end{pmatrix}.
\end{align}
The TR-symmetry of the system induces the following relation on Wilson loop operator \cite{PhysRevB.96.245115}
\begin{align} \label{eq:W_symm}
    w(k_x,k_y)W^*_{\mathbf{e}_2,k_y}(k_x)w(k_x,k_y)^{-1} = W^{-1}_{\mathbf{e}_2,-k_y}(-k_x).
\end{align}
Since the Wilson loop operator in this system is real due to the composite $T C_2$ symmetry, $W^*$ in Eq. (\ref{eq:W_symm}) can be taken to be $W$ . Let the start point of the Wilson loop be $(k_x,k_y) = (k_x,\pi)$. Then the sewing matrix $w$ at the $k_y=\pi$ line can then be shown to be independent of $k_x$ as in appendix \ref{sec: sewing},
\begin{align} \label{eq: Wilson_T_symm}
    w(k_x,\pi) & = q_0(k_x) \sigma_z \otimes \sigma_0 \notag \\
    & = 
    \begin{pmatrix}
        0 & -I_{2 \times 2} \\
        I_{2 \times 2} & 0
    \end{pmatrix}.
\end{align}
Eq. (\ref{eq:W_symm}) becomes
\begin{align} \label{eq: W_Tsymm}
    w W_{\mathbf{e}_2,\pi}(k_x)w^{-1} = W^{-1}_{\mathbf{e}_2,\pi}(-k_x),
\end{align}
where $w=\begin{pmatrix}
    0 & -I_{2 \times 2} \\
    I_{2 \times 2} & 0
\end{pmatrix}$.
Note the Wilson loop $W_{\mathbf{e}_2,k_y=\pi}(k_x)$ (\ref{eq: wil}) and the sewing matrix $w$ of TR-symmetry in this example have been analysed in the previous section. The new topological invariant $\nu_{new}$ is equal to $n \mod 2$. For odd $n$, we have shown in the previous section that Wannier bands in its Wilson loop spectrum can not unwind. 

\begin{figure} 
    \centering
    \includegraphics[width=.8\columnwidth]{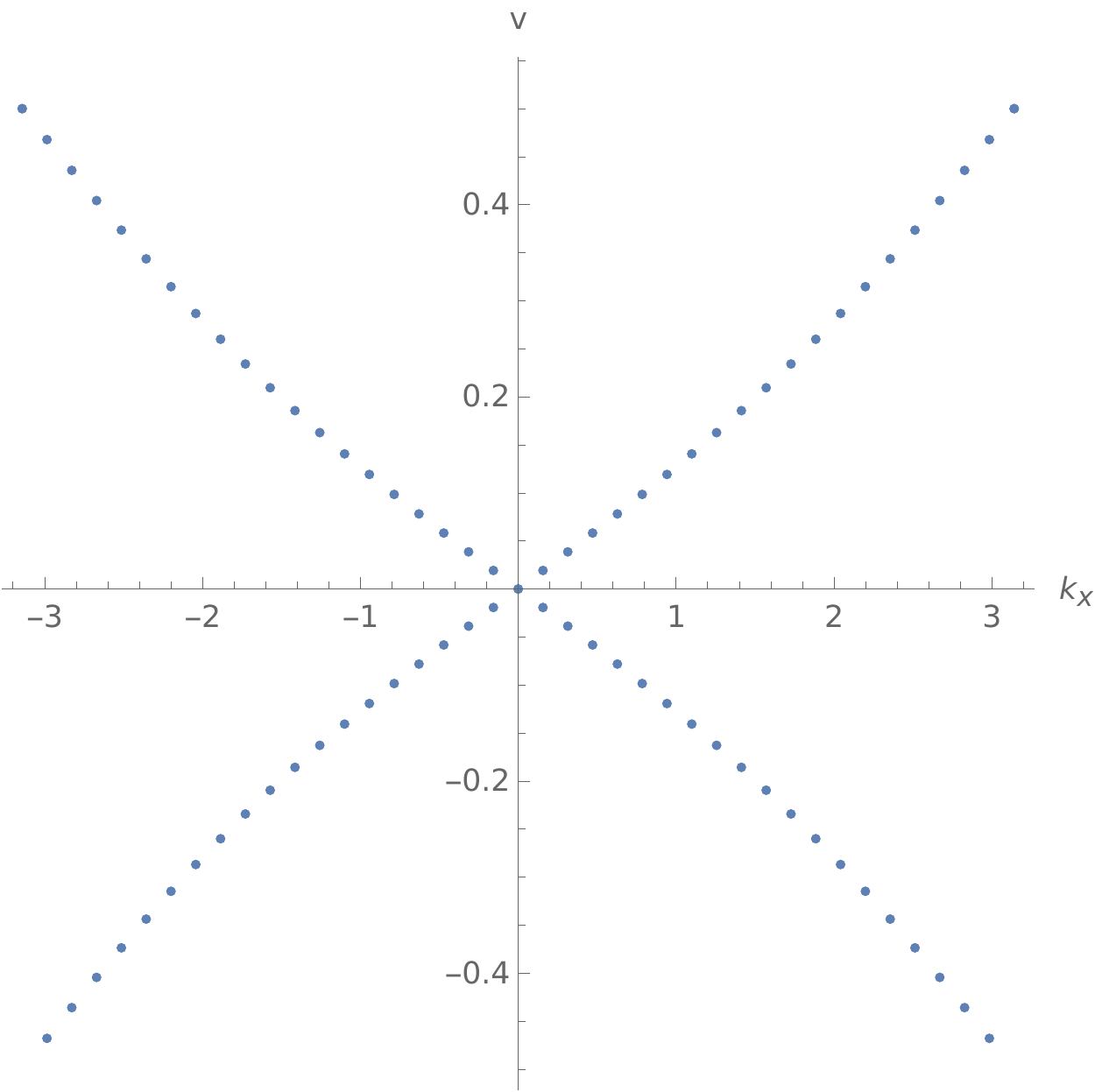}
    \caption[Spectrum]{The Wilson loop spectrum of the tight binding model. Each band has a twofold degenercy.} \label{fig: spectrum}
\end{figure}

We apply a cutoff on the Fourier transformation of the Hamiltonian (\ref{eq:ham}) with $n=1$ in momentum space to obtain a tight binding model. The hopping matrices are taken to be 
\begin{align}
    t_{00} & = \left(
        \begin{array}{cccccccc}
         0 & 0 & 0 & 0 & 0 & 0 & \frac{1}{\pi } & 0 \\
         0 & 0 & 0 & 0 & 0 & 0 & 0 & \frac{1}{\pi } \\
         0 & 0 & 0 & 0 & \frac{1}{\pi } & 0 & 0 & 0 \\
         0 & 0 & 0 & 0 & 0 & \frac{1}{\pi } & 0 & 0 \\
         0 & 0 & \frac{1}{\pi } & 0 & 0 & 0 & 0 & 0 \\
         0 & 0 & 0 & \frac{1}{\pi } & 0 & 0 & 0 & 0 \\
         \frac{1}{\pi } & 0 & 0 & 0 & 0 & 0 & 0 & 0 \\
         0 & \frac{1}{\pi } & 0 & 0 & 0 & 0 & 0 & 0 \\
        \end{array}
        \right) , \notag 
\end{align}
\begin{align}
    t_{01} & = \left(
        \begin{array}{cccccccc}
         -\frac{i}{2} & 0 & 0 & 0 & 0 & 0 & -\frac{1}{4} & 0 \\
         0 & -\frac{i}{2} & 0 & 0 & 0 & 0 & 0 & -\frac{1}{4} \\
         0 & 0 & -\frac{i}{2} & 0 & -\frac{1}{4} & 0 & 0 & 0 \\
         0 & 0 & 0 & -\frac{i}{2} & 0 & -\frac{1}{4} & 0 & 0 \\
         0 & 0 & -\frac{1}{4} & 0 & \frac{i}{2} & 0 & 0 & 0 \\
         0 & 0 & 0 & -\frac{1}{4} & 0 & \frac{i}{2} & 0 & 0 \\
         -\frac{1}{4} & 0 & 0 & 0 & 0 & 0 & \frac{i}{2} & 0 \\
         0 & -\frac{1}{4} & 0 & 0 & 0 & 0 & 0 & \frac{i}{2} \\
        \end{array}
        \right) , \notag
\end{align}
\begin{align}
    t_{02} & = \left(
        \begin{array}{cccccccc}
         0 & 0 & 0 & 0 & 0 & 0 & \frac{1}{3 \pi } & 0 \\
         0 & 0 & 0 & 0 & 0 & 0 & 0 & \frac{1}{3 \pi } \\
         0 & 0 & 0 & 0 & \frac{1}{3 \pi } & 0 & 0 & 0 \\
         0 & 0 & 0 & 0 & 0 & \frac{1}{3 \pi } & 0 & 0 \\
         0 & 0 & \frac{1}{3 \pi } & 0 & 0 & 0 & 0 & 0 \\
         0 & 0 & 0 & \frac{1}{3 \pi } & 0 & 0 & 0 & 0 \\
         \frac{1}{3 \pi } & 0 & 0 & 0 & 0 & 0 & 0 & 0 \\
         0 & \frac{1}{3 \pi } & 0 & 0 & 0 & 0 & 0 & 0 \\
        \end{array}
        \right) , \notag
\end{align}
\begin{align}
    t_{10} & = \left(
        \begin{array}{cccccccc}
         0 & 0 & 0 & 0 & 0 & 0 & \frac{1}{2 \pi } & \frac{i}{2 \pi } \\
         0 & 0 & 0 & 0 & 0 & 0 & -\frac{i}{2 \pi } & \frac{1}{2 \pi } \\
         0 & 0 & 0 & 0 & \frac{1}{2 \pi } & \frac{i}{2 \pi } & 0 & 0 \\
         0 & 0 & 0 & 0 & -\frac{i}{2 \pi } & \frac{1}{2 \pi } & 0 & 0 \\
         0 & 0 & \frac{1}{2 \pi } & -\frac{i}{2 \pi } & 0 & 0 & 0 & 0 \\
         0 & 0 & \frac{i}{2 \pi } & \frac{1}{2 \pi } & 0 & 0 & 0 & 0 \\
         \frac{1}{2 \pi } & -\frac{i}{2 \pi } & 0 & 0 & 0 & 0 & 0 & 0 \\
         \frac{i}{2 \pi } & \frac{1}{2 \pi } & 0 & 0 & 0 & 0 & 0 & 0 \\
        \end{array}
        \right) , \notag 
\end{align}
\begin{align}
    t_{11} & = \left(
        \begin{array}{cccccccc}
         0 & 0 & 0 & 0 & 0 & 0 & \frac{1}{8} & \frac{i}{8} \\
         0 & 0 & 0 & 0 & 0 & 0 & -\frac{i}{8} & \frac{1}{8} \\
         0 & 0 & 0 & 0 & \frac{1}{8} & \frac{i}{8} & 0 & 0 \\
         0 & 0 & 0 & 0 & -\frac{i}{8} & \frac{1}{8} & 0 & 0 \\
         0 & 0 & \frac{1}{8} & -\frac{i}{8} & 0 & 0 & 0 & 0 \\
         0 & 0 & \frac{i}{8} & \frac{1}{8} & 0 & 0 & 0 & 0 \\
         \frac{1}{8} & -\frac{i}{8} & 0 & 0 & 0 & 0 & 0 & 0 \\
         \frac{i}{8} & \frac{1}{8} & 0 & 0 & 0 & 0 & 0 & 0 \\
        \end{array}
        \right) , \notag 
\end{align}
\begin{align}
    t_{12} & = \left(
        \begin{array}{cccccccc}
         0 & 0 & 0 & 0 & 0 & 0 & \frac{1}{6 \pi } & \frac{i}{6 \pi } \\
         0 & 0 & 0 & 0 & 0 & 0 & -\frac{i}{6 \pi } & \frac{1}{6 \pi } \\
         0 & 0 & 0 & 0 & \frac{1}{6 \pi } & \frac{i}{6 \pi } & 0 & 0 \\
         0 & 0 & 0 & 0 & -\frac{i}{6 \pi } & \frac{1}{6 \pi } & 0 & 0 \\
         0 & 0 & \frac{1}{6 \pi } & -\frac{i}{6 \pi } & 0 & 0 & 0 & 0 \\
         0 & 0 & \frac{i}{6 \pi } & \frac{1}{6 \pi } & 0 & 0 & 0 & 0 \\
         \frac{1}{6 \pi } & -\frac{i}{6 \pi } & 0 & 0 & 0 & 0 & 0 & 0 \\
         \frac{i}{6 \pi } & \frac{1}{6 \pi } & 0 & 0 & 0 & 0 & 0 & 0 \\is
        \end{array}
        \right),
\end{align}
where $t_{ij}$ represents the hopping parameter that an electron hops from site $(m,n)$ to site $(m+i,n+j)$ and it is a $8 \times 8$-matrix since every unit cell has eight internal degrees of freedom. Further, hopping matrices in the inverse direction are taken to be conjugate transpositions of these tabulated hopping matrices, and any other hopping matrices are zero. The Wilson loop spectrum of this system is plotted in figure \ref{fig: spectrum}. In Table IV of Ref. \cite{PhysRevB.100.115160}, possible atomic insulating states are listed which all have a gapped Wilson loop spectrum. Hence, the gapless Wilson loop spectrum presents an obstruction to the Wannier representation of this phase, and it implies that this phase is in a fragile topological insulating phase or in a stable band topology phase \cite{PhysRevLett.121.126402}. We believe that this phase is in a stable band topology phase since the topological invariant $\nu_{new}$ is stable.

\section{Conclusions} \label{sec: conclusion}
In this work we present a new homotopy invariant in twofold rotation symmetric system in AZ class AII. We prove that it agrees with the K theory topological invariant. The main idea is lifting the Wilson loop to the universal covering group, and the topology origin of this invariant is the disconnectedness of the fixed point set of TR-symmetry acting on the covering group. We show two examples belonging to two symmetry categories in a case of four occupied bands that have the same Wilson loop spectrum but belong to different topological classes (one is trivial and another is non-trivial).

In addition, we have shown in the case of four occupied bands that even when three other topological invariants(partial polarizations and the Fu-Kane-Mele invariant) vanish, there is an additional obstruction that prevents the Wilson loop spectrum from unwinding. This extends the results in Refs. \cite{PhysRevB.99.045140,PhysRevB.100.195135}, in which the conservation of Wilson loop spectrum winding number in the case of two occupied bands and the unwinding of the Wilson loop spectrum in the case of more occupied bands are studied.

\begin{acknowledgments}
    The authors thank Yongxu Fu, Shuxuan Wang for discussions. This work was supported by NSFC Grant No.11275180.
\end{acknowledgments}

\appendix

\section{Computation of the K group} \label{sec: K group}
\begin{figure}
    \centering
    \subfigure[]{
        \centering
        \includegraphics[width=.3\columnwidth]{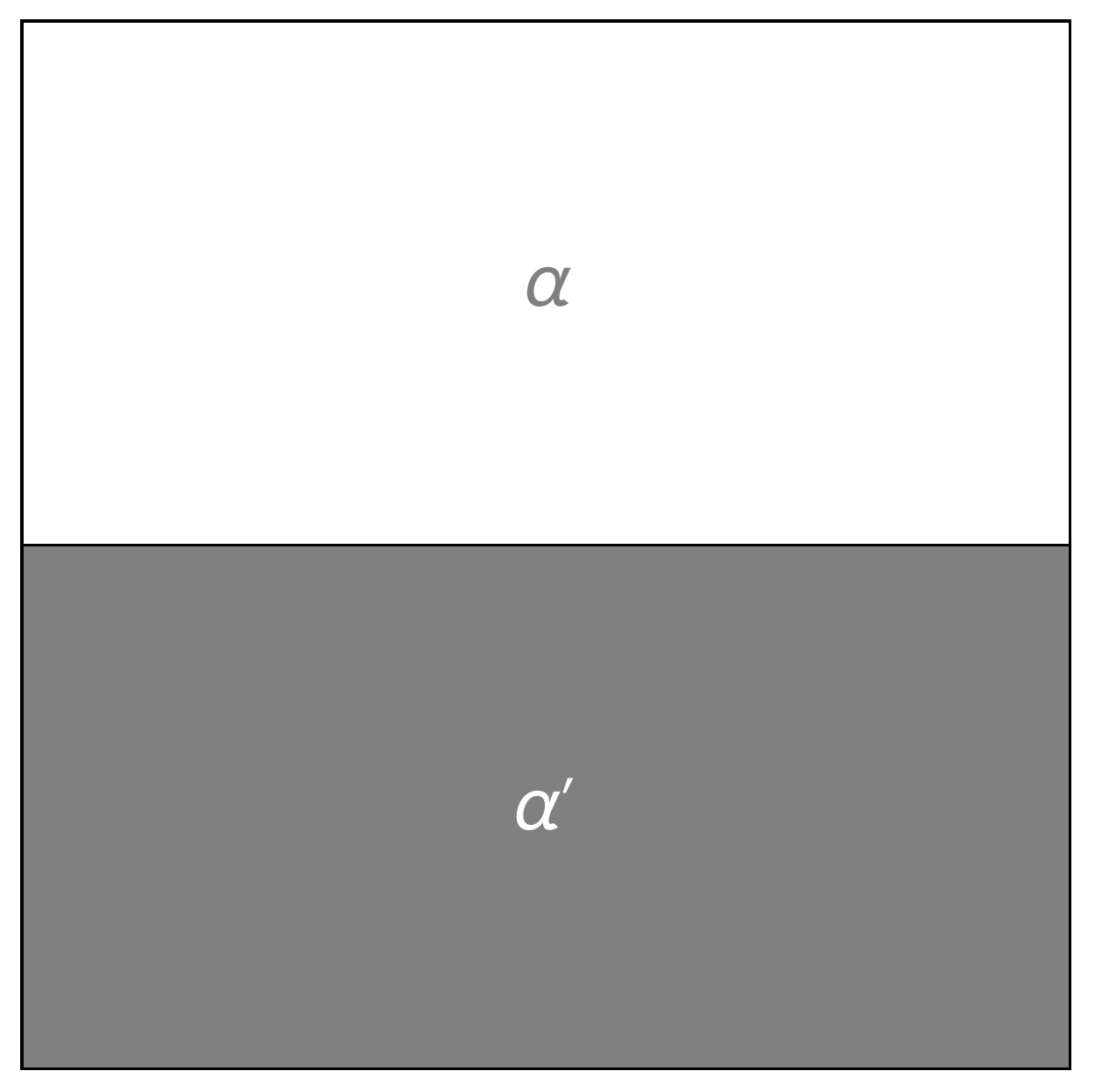}
    } 
    \subfigure[]{
        \centering
        \includegraphics[width=.3\columnwidth]{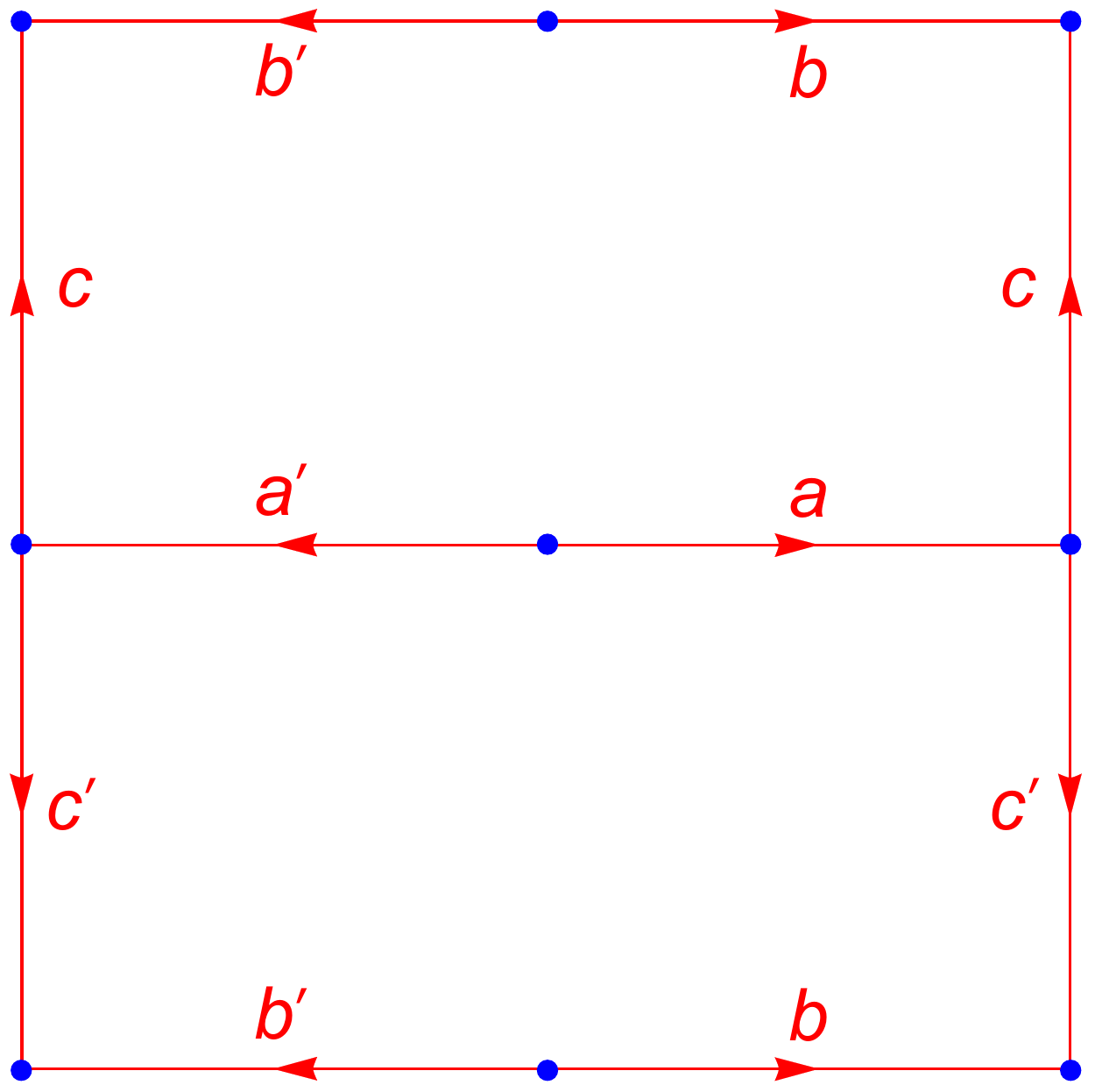}
    }
    \subfigure[]{
        \centering
        \includegraphics[width=.3\columnwidth]{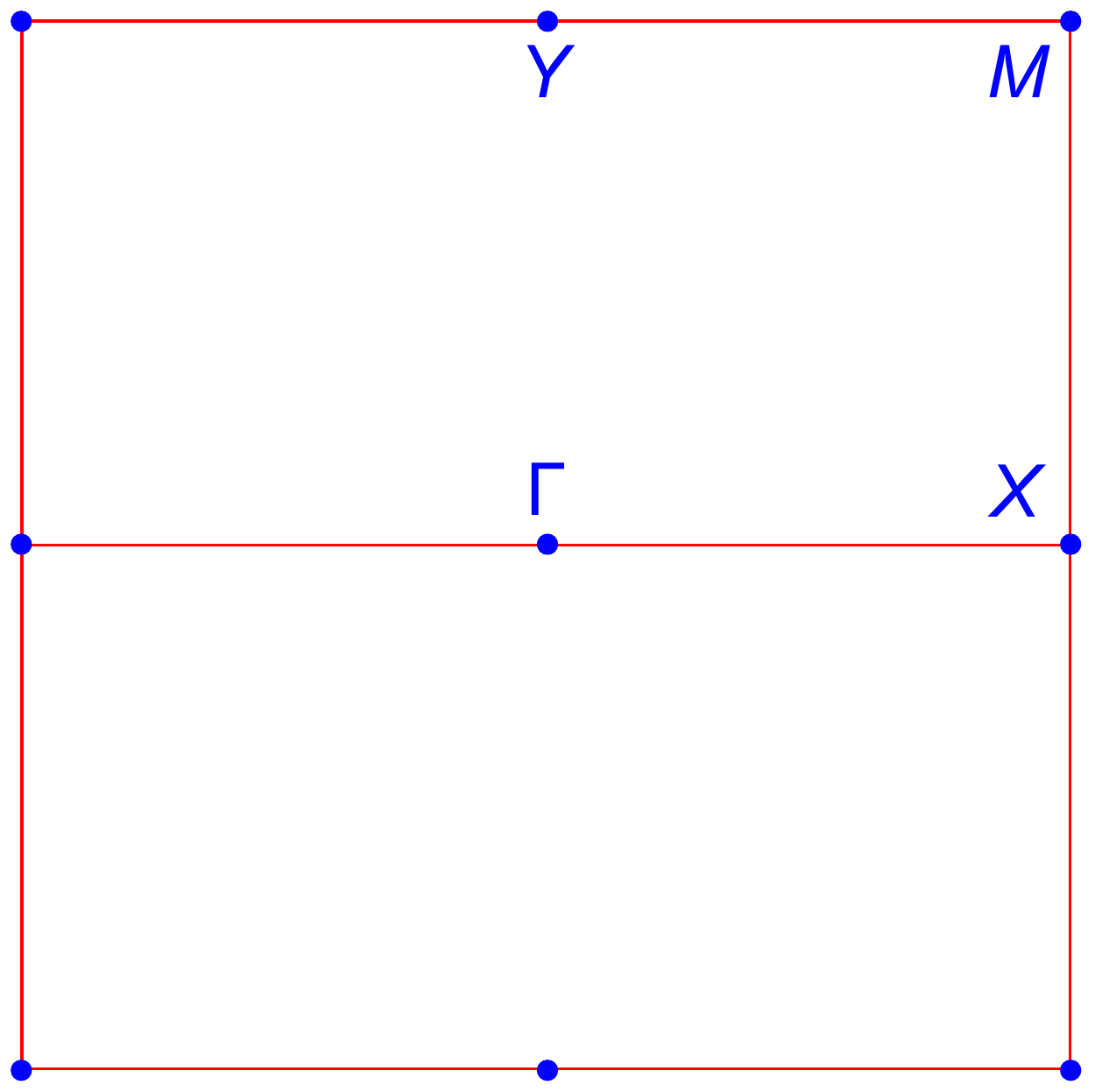}
    } 
    \caption[Cell decomposition of the Brillouin zone]{Cell decomposition of the Brillouin zone $\mathbb{T}^2$. (a) Two dimensional cells $\alpha, \alpha '$ are shown. (b) One dimensional cells $a,b,c,a',b',c'$ are shown as red arrows. (c) Zero dimensional cells $\Gamma, X, Y, M$ are shown as blue points. } \label{fig: AppeCell}
\end{figure}

In this appendix, we present a calculation of the twisted equivariant K group $\prescript{\phi}{}{K}^{(\tau,c)}_G(X)$ of the system using the spectral sequence method \cite{shiozaki2018atiyahhirzebruch}. Another approach using dimension raising isomorphism also works \cite{PhysRevB.82.115120,PhysRevB.90.165114,PhysRevB.95.235425}. Here we prefer the spectral sequence method for the convenience of relating topological invariants to pages in a spectral sequence. We simply give the calculation here; the details and an explanation of this method can be found in Ref. \cite{shiozaki2018atiyahhirzebruch}.  The first step of this method is processing an equivariant cell decomposition, which is shown in Fig. \ref{fig: AppeCell}.

The symmetry group of each zero dimensional cell is $G=\mathbb{Z}_2^{C_2} \times \mathbb{Z}_2^T$. The little group $G$ splits into the disjoint union of left cosets as
\begin{align}
    G = G^0 \sqcup T G^0,
\end{align}
where $G^0 = \{ g \in G | \phi(g) = c(g) = 1 \} = \mathbb{Z}_2^{C_2}$ is the subgroup of unitary symmetries. Time-reversal symmetry $T \in G$ is a magnetic symmetry. There are two twisted irreducible representations of $G^0= \mathbb{Z}_2^{C_2}$: the one dimensional representation with $C_2=i$, and the one dimensional representation with $C_2=-i$. For two such irreducible representations, the Wigner test can be applied, with which we can calculate the following formula
\begin{align}
    W^T_\alpha = \frac{1}{|G^0|} \sum_{g \in G^0} z_{ag,ag} \chi_{\alpha} ((a g)^2),
\end{align}
where $\alpha$ denotes the corresponding irreducible representation, $z$ is the factor system of $G$ and $\chi_{\alpha}$ is the character of the irreducible representation $\alpha$. For the above two irreducible representations, we obtain
\begin{align}
    W^T_{C_2 = i} & = 0 \notag \\
    W^T_{C_2 = -i} & = 0,
\end{align}
which implies that the symmetry classes of these zero dimensional cells are AZ class A, and these two irreducible representations are tied by the magnetic symmetry $T$. We can apply the same process for one dimensional and two dimensional cells. Each cell has symmetry group $G_k = \{ e, TC_2 \}$, the subgroup of unitary symmetries $G_k^0 = \{e\}$ and the magnetic symmetry $a= TC_2$. The result of the Wigner test is $W^T_{\mathrm{trivial}} = 1$ which implies their symmetry classes are AZ class AI. This can be intuitively understood since $TC_2$ is a magnetic symmetry that satisfies $(TC_2)^2=1$. Use the above facts, the first page of the spectral sequence $E_1^{p,-n} = \prod_{j \in I^p_{orb}} \prescript{\phi}{}{K}^{(\tau,c),-(n-p)}_G(X_p,X_{p-1}) = \prod_{j \in I^p_{orb}} \prescript{\phi|_{D_j^p}}{}{K}^{(\tau,c)|_{D^p_j} , -n}_{G_{D^p_j}}(D^p_j)$ is listed in Table \ref{tab:E1page}.
\begin{table} \centering
    \begin{tabular}{c | c  c  c}
        \hline \hline
        $E^{p,-n}_1$ & $\{\Gamma,X,Y,M\}$ & $\{a,b,c\}$ & $\{\alpha\}$ \tabularnewline \hline
        n=0 & $\mathbb{Z}+\mathbb{Z}+\mathbb{Z}+\mathbb{Z}$ & $\mathbb{Z}+\mathbb{Z}+\mathbb{Z}$ & $\mathbb{Z}$ \tabularnewline
        n=1 & 0 & $\mathbb{Z}_2+\mathbb{Z}_2+\mathbb{Z}_2$ & $\mathbb{Z}_2$ \tabularnewline
        n=2 & $\mathbb{Z}+\mathbb{Z}+\mathbb{Z}+\mathbb{Z}$ & $\mathbb{Z}_2+\mathbb{Z}_2+\mathbb{Z}_2$ & $\mathbb{Z}_2$ \tabularnewline
        n=3 & 0 & 0 & 0 \tabularnewline
        \hline \hline
    \end{tabular}
    \caption{The $E_1$ page of the spectral sequence.} \label{tab:E1page}
\end{table}

The only non-vanishing differential for $n \in \{0,1,2,3\}$ is $d_1^{0,0} : E_1^{0,0} \rightarrow E_1^{1,0}$, which is the set of compatible relations of symmetry indicators at high symmetry points. The value of $d_1^{0,0} : E_1^{0,0} \rightarrow E_1^{1,0}$ is listed in Table \ref{tab:differential}.
\begin{table*} \centering
    \begin{tabular}{c c c c | c}
        \hline \hline
        $\Gamma_{\left[ C2=\begin{pmatrix} i & 0 \\ 0 & -i \end{pmatrix} \right]}$ & $X_{\left[ C2=\begin{pmatrix} i & 0 \\ 0 & -i \end{pmatrix} \right]}$ & $Y_{\left[ C2=\begin{pmatrix} i & 0 \\ 0 & -i \end{pmatrix} \right]}$ & $M_{\left[ C2=\begin{pmatrix} i & 0 \\ 0 & -i \end{pmatrix} \right]}$ & \tabularnewline 
        &&&& \tabularnewline \hline
        2 & -2 & 0 & 0 & $a$ \tabularnewline
        0 & 2 & 0 & -2 & $b$ \tabularnewline
        0 & 0 & 2 & -2 & $c$ \tabularnewline
        \hline \hline
    \end{tabular}
    \caption{The differential $d_1^{0,0} : E_1^{0,0} \rightarrow E_1^{1,0}$. The subscript $C_2=\begin{pmatrix} i & 0 \\ 0 & -i \end{pmatrix}$ under each high symmetry point symbol means the irreducible representations $C_2=i$ and $C_2=-i$ come in pair due to time-reversal symmetry.} \label{tab:differential}
\end{table*}

One can calculate the $E_2$ page of the spectral sequence from the above data via the definition
\begin{align}
    E^{p,-n}_2 = \ker d^{p,-n}_1 / \im d^{p-1,-n}_1 .
\end{align}
The $E_2$ page is listed in Table \ref{tab:E2page}.
\begin{table} \centering
    \begin{tabular}{c | c  c  c}
        \hline \hline
        $E^{p,-n}_2$ & $p=0$ & $p=1$ & $p=2$ \tabularnewline \hline
        n=0 & $\mathbb{Z}$ & $\mathbb{Z}_2+\mathbb{Z}_2+\mathbb{Z}_2$ & $\mathbb{Z}$ \tabularnewline
        n=1 & 0 & $\mathbb{Z}_2+\mathbb{Z}_2+\mathbb{Z}_2$ & $\mathbb{Z}_2$ \tabularnewline
        n=2 & $\mathbb{Z}+\mathbb{Z}+\mathbb{Z}+\mathbb{Z}$ & $\mathbb{Z}_2+\mathbb{Z}_2+\mathbb{Z}_2$ & $\mathbb{Z}_2$ \tabularnewline
        n=3 & 0 & 0 & 0 \tabularnewline
        \hline \hline
    \end{tabular}
    \caption{The $E_2$ page of the spectral sequence.} \label{tab:E2page}
\end{table}
One can see that all the second order differentials $d^{p,-n}_2 : E^{p,-n}_2 \rightarrow E^{p+2,-(n+1)}_2$ for $n<3$ vanish. The third page $E_3$ of the spectral sequence which is also the infinite page $E_{\infty}$, is listed in Table \ref{tab:E3page}.
\begin{table} \centering
    \begin{tabular}{c | c  c  c}
        \hline \hline
        $E^{p,-n}_3$ & $p=0$ & $p=1$ & $p=2$ \tabularnewline \hline
        n=0 & $\mathbb{Z}$ & $\mathbb{Z}_2+\mathbb{Z}_2+\mathbb{Z}_2$ & $\mathbb{Z}$ \tabularnewline
        n=1 & 0 & $\mathbb{Z}_2+\mathbb{Z}_2+\mathbb{Z}_2$ & $\mathbb{Z}_2$ \tabularnewline
        n=2 & $\mathbb{Z}+\mathbb{Z}+\mathbb{Z}+\mathbb{Z}$ & $\mathbb{Z}_2+\mathbb{Z}_2+\mathbb{Z}_2$ & $\mathbb{Z}_2$ \tabularnewline
        \hline \hline
    \end{tabular}
    \caption{The $E_3=E_{\infty}$ page of the spectral sequence.} \label{tab:E3page}
\end{table}

From these pages of the spectral sequence, we conclude
\begin{align}
    E^{0,0}_{\infty} & = E^{0,0}_2 \subseteq \prescript{\phi|_{X_0}}{}{K}^{(\tau,c)|_{X_0}}_G(X_0) \notag \\
    E^{1,-1}_{\infty} & = E^{1,-1}_1 = \prescript{\phi|_{X_1}}{}{K}^{(\tau,c)|_{X_1}}_G(X_1,X_0) \notag \\
    E^{2,-2}_{\infty} & = E^{2,-2}_1 = \prescript{\phi}{}{K}^{(\tau,c)}_G(X_2,X_1),
\end{align}
where the first term $E^{0,0}_2 \cong \mathbb{Z}$ is characterized by the filling number of the system, the second term $E^{1,-1}_{\infty} = \prescript{\phi|_{X_1}}{}{K}^{(\tau,c)|_{X_1}}_G(X_1,X_0) \cong \mathbb{Z}_2+\mathbb{Z}_2+\mathbb{Z}_2$ is characterized by three $\mathbb{Z}_2$-valued topological invariants which are $\nu_{\Gamma X}$,$\nu_{\Gamma Y}$ and $\nu_{\mathrm{FKM}}$, and the third term $E^{2,-2}_{\infty} = \prescript{\phi}{}{K}^{(\tau,c)}_G(X_2,X_1) \cong \mathbb{Z}_2$ is characterized by a new $\mathbb{Z}_2$-valued topological invariant $\nu_{new}$. The K group of the system $\prescript{\phi}{}{K}^{(\tau,c)}_G(X)$ relates to the infinite pages of the spectral sequence by following two short exact sequences,
\begin{align}
    0 \rightarrow F^{1,-1} \rightarrow \prescript{\phi}{}{K}^{(\tau,c)}_G(X) \rightarrow E^{0,0}_{\infty} \rightarrow 0, \notag \\
    0 \rightarrow E^{2,-2}_{\infty} \rightarrow F^{1,-1} \rightarrow E^{1,-1}_{\infty} \rightarrow 0.
\end{align}
Since $E^{0,0}_{\infty} \cong \mathbb{Z}$ is a free $\mathbb{Z}$-module, the first short exact sequence splits and we obtain
\begin{align}
    \prescript{\phi}{}{K}^{(\tau,c)}_G(X) \cong E^{0,0}_{\infty} \oplus F^{1,-1} \cong \mathbb{Z} \oplus F^{1,-1}.
\end{align}
By the dimension raising isomorphism argument \cite{PhysRevB.90.165114}, we already know the reduced version of K group $\prescript{\phi}{}{\tilde{K}}^{(\tau,c)}_G(X) \cong \mathbb{Z}_2^4$, which requires that the second short exact sequence also splits,
\begin{align}
    F^{1,-1} \cong E^{2,-2}_{\infty} \oplus E^{1,-1}_{\infty} \cong \mathbb{Z}_2^4.
\end{align}

In conclusion,
\begin{align}
    \prescript{\phi}{}{K}^{(\tau,c)}_G(X) \cong E^{0,0}_2 & \oplus \prescript{\phi|_{X_1}}{}{K}^{(\tau,c)|_{X_1}}_G(X_1,X_0) \notag \\
    & \oplus \prescript{\phi}{}{K}^{(\tau,c)}_G(X_2,X_1),
\end{align}
where the first component $E^{0,0}_2 \cong \mathbb{Z}$ is characterized by the filling number of the system, the second component $\prescript{\phi|_{X_1}}{}{K}^{(\tau,c)|_{X_1}}_G(X_1,X_0)$ is characterized by three $\mathbb{Z}_2$-valued topological invariants, which are $\nu_{\Gamma X}$,$\nu_{\Gamma Y}$ and $\nu_{\mathrm{FKM}}$, and the third component $\prescript{\phi}{}{K}^{(\tau,c)}_G(X_2,X_1)$ is characterized by a new $\mathbb{Z}_2$-valued topological invariant $\nu_{new}$, which is focused on in this paper.

\section{Proof of theorem \ref{thm: trivial}} \label{sec: proof}
\begin{proof}
    
    We divide the proof into two parts. The first part assume that each lifted Wilson loop $\tilde{W}_{\mathbf{e}_2,\pi}(k_x)$ has a base point $\tilde{W}_{\mathbf{e}_2,\pi}(k_x=0) = (1,1)$. The homotopy in this part should fix this base point and preserve all symmetries. The second part of the proof allow a flowing of the base point, i.e. the homotopy without a base point is studied. In this part, we show that the two lifted Wilson loops with base points $(1,1)$ and $(1,h) \in X_0$ are topologically equivalent, where $X_0 = \{ (1,x'_0+x'_1 i + x'_3 k) | (x'_0)^2 + (x'_1)^2 + (x'_3)^2 = 1\}$ as shown in the main text.

    Now, we show the first part of the proof. First, we prove that if $\tilde{W}_{\mathbf{e}_2,\pi}(k_x = \pi) \in Y_0$, the system is non-trivial. Assume that a homotopy between $k_x \mapsto W_{\mathbf{e}_2,\pi}(k_x)$ and $k_x \mapsto I_{4 \times 4}$ exists, and denote it by $W(k_x,t)$. Since the midpoint of the lifted Wilson loop is located at the fixed point set $X_0 \bigsqcup Y_0$, we have a continuous path $\tilde{W}(\pi,t) \in X_0 \bigsqcup Y_0$. Its start point is $\tilde{W}(\pi,0) \in Y_0$, and its end point is $\tilde{W}(\pi,1) \in X_0$. Since $X_0$ and $Y_0$ are disconnected in $Sp(1) \times Sp(1)$, they leads to a contradiction.

    Secondly, we prove that if $\tilde{W}_{\mathbf{e}_2,\pi}(k_x = \pi) \in X_0$, the system is trivial. It is sufficient to consider the half path of the lifting $k_x \mapsto \tilde{W}_{\mathbf{e}_2,\pi}(k_x), k_x \in [0, \pi]$. The other half can be constructed via the time-reversal operator $T$,
    \begin{align}
        \tilde{W}_{\mathbf{e}_2,\pi}(2 \pi - k_x) =T(\tilde{W}_{\mathbf{e}_2,\pi}(k_x)).
    \end{align}
    The topological classification of the half path of the lifting $k_x \mapsto \tilde{W}_{\mathbf{e}_2,\pi}(k_x), k_x \in [0, \pi]$ is given by the relative homotopy group $\pi_1(Sp(1) \times Sp(1),X_0)$, which is trivial as seen in the following exact sequence,
    \begin{align}
        (\pi_1(X_0)=\pi_1(S^2) \cong 0) & \rightarrow (\pi_1(Sp(1) \times Sp(1)) \cong 0) \notag \\
        & \rightarrow \pi_1(Sp(1) \times Sp(1),X_0) \notag \\
        & \rightarrow (\pi_0(X_0) \cong 0).
    \end{align}
    Hence any lifting path $k_x \mapsto \tilde{W}_{\mathbf{e}_2,\pi}(k_x)$ with $\tilde{W}_{\mathbf{e}_2,\pi}(k_x = \pi) \in X_0$ is homotopic to a constant path $k_x \mapsto (1,1)$ which is trivial. It follows that $k_x \mapsto W_{\mathbf{e}_2,\pi}(k_x)$ is homotopic to a constant loop $k_x \mapsto I_{4 \times 4}$ with respect to symmetries.

    We show the second part of the proof by considering a lifted Wilson loop $\tilde{W}_{\mathbf{e}_2,\pi}(k_x)$ with a base point $(1,1)$ and flows it to any base point $(1,h) \in X_0$ via a homotopy preserving all symmetries. Since there is a connected path $p(t),t \in [0,1]$ with the start point $(1,1)$ and the end point $(1,h)$ in $X_0$, we explicitly construct this homotopy as
    \begin{align}
        \tilde{W}(k,t) = p(t) \cdot \tilde{W}_{\mathbf{e}_2,\pi}(k) .
    \end{align}
    In this setting, TR symmetry is preserved, as shown by
    \begin{align}
        & \tilde{w}^{-1} \tilde{W}(k,t) \tilde{w} \notag \\
        = & (1,-j) \cdot p(t) \cdot \tilde{W}_{\mathbf{e}_2,\pi}(k) \cdot (1,j) \notag \\
        = & \overline{p(t)} (1,-j) \cdot \tilde{W}_{\mathbf{e}_2,\pi}(k) \cdot (1,j) \notag \\
        = & \overline{p(t)} \cdot \overline{\tilde{W}_{\mathbf{e}_2,\pi}(-k)}  \\
        = & \overline{p(t) \tilde{W}_{\mathbf{e}_2,\pi}(-k)} \notag \\
        = & \overline{\tilde{W}(-k,t)} \notag  ,
    \end{align}
    where in the second equality the relation $(1,-j) \cdot a = \overline{a} \cdot (1,-j)$ for any $a \in X_0$ is used.

    Furthermore, the base point of the final lifted Wilson loop becomes $\tilde{W}(0,t=1) = (1,h) \cdot \tilde{W}_{\mathbf{e}_2,\pi}(k=0) = (1,h)$.
\end{proof}

\section{Canonical form of Wilson loops} \label{sec: can_form}
In section \ref{sec: wil_examples} of the main text, we give two examples of Wilson loops which are block diagonal. In this appendix, we denote each of them as the canonical form of Wilson loops in each category. We give a proof on the statement under the assumption that we consider a system with a vanishing partial polarization invariant $\nu_{\mathbf{\Gamma Y}}$ and a vanishing FKM invariant, and that any Wilson loop of such a system is homotopic equivalent to the canonical form with respect to all symmetries. Furthermore, we give an explicit example of a deformation from a gapped Wilson loop to a gapless Wilson loop in the canonical form.

The proof of the existence of a symmetric deformation to the canonical form can be simply given by using our proved theorem \ref{thm: trivial}. Consider the vanishing partial polarization invariant $\nu_{\mathbf{\Gamma Y}}$ and the vanishing FKM invariant case. By theorem \ref{thm: trivial}, any Wilson loop can be deformed to representives of homotopy equivalent classes with respect to all symmetries. In this case, there are two homotopy equivalent classes, one that satisfies $\nu_{new}=0 \mod 2$ and another that satisfies $\nu_{new}=1 \mod 2$. In each symmetry category, these two representives can be found in the set of Wilson loops in canonical form. To avoid confusion, we emphasize here that the first category actually contains both representives, hence, to prove the equivalence of our homotopy invariant and the K theory invariant, we only need examples belonging to the first category(or the second category). The reason that we introduce both categories is that it provides more examples and shows that systems with different homotopy invariant $\nu_{new}$ may have the same Wilson loop spectrum.

Here we give an explicit example of a deformation from a gapped Wilson loop to a gapless Wilson loop in the canonical form. Consider a Wilson loop and a sewing matrix of TR symmetry of the following forms: 
\begin{align}
    W & = \mathbf{diag}(A,B), \notag \\
    w & = \begin{pmatrix}
        0 & 0 & -1 & 0 \\
        0 & 0 & 0 & -1 \\
        1 & 0 & 0 & 0 \\
        0 & 1 & 0 & 0
    \end{pmatrix} , \notag
\end{align}
where $A= \begin{pmatrix}
    \cos (m k + \alpha)& \sin (m k + \alpha) \\
    -\sin (m k + \alpha) & \cos (m k + \alpha)
\end{pmatrix}$ and $B = \begin{pmatrix}
    \cos (m k - \alpha) & \sin (m k - \alpha) \\
    -\sin (m k - \alpha) & \cos (m k - \alpha)
\end{pmatrix}$.

Note that the above Wilson loop is TR invariant for any $\alpha$. By a linear interpolation, this Wilson loop is homotopic to $\mathbf{diag}(A,B)$ with $\alpha=0$ which is the canonical form in section \ref{sec: wil_examples} of the main text. Also note that when $m=0$ and $\alpha \neq 0,\pi$, the Wilson loop spectrum(Wannier bands) is gapped. Use the Hamiltonian (\ref{eq:ham}) in the main text, and set $q_1(k_x)$ as
\begin{align}
    q_1(k_x) =
    \begin{pmatrix}
        0 & B^T \\
        A^T & 0
    \end{pmatrix},
\end{align}
where $A$ and $B$ are same as above. In the deformation process in which $\alpha$ changes from its initial non-zero value to zero, the Wannier band gap is closed while the gap of the Hamiltonian (\ref{eq:ham}) remains open. Hence, the Wannier gap is not topologically protected in this case, and this gapped Wilson loop is deformed to the canonical form with repect to all symmetries.

Now consider a system with a non-zero partial polarization invariant $\nu_{\mathbf{\Gamma Y}}$. Its Wilson loop is
\begin{align}
    W = 
    \begin{pmatrix}
        \cos m k & \sin m k & 0 & 0 \\
        -\sin m k & \cos m k & 0 & 0 \\
        0 & 0 & -\cos n k & -\sin n k \\
        0 & 0 & \sin n k & -\cos n k
    \end{pmatrix} ,
\end{align}
and the Wilson loop with $m=n=0$ is gapped. Here the Wannier band gap is still not topologically protected since the $m=n=0$ phase is connected to the $m=n=2$ phase, which is gapless, without closing the gap of the Hamiltonian and with respect to all symmetries.

\section{Computation of the Wilson loop} \label{sec: Wilson}
In this section, we show two facts used in the main text. Firstly, we compute the Wilson loop matrix of the example mentioned in the main text. We show that it can be analytically computed. Furthermore, we provide an explicit inverse map of the dimension raising map. Hence we show the second fact, namely that the classification problem of the Hamiltonian with respect to symmetries is equivalent to the classification problem of the path of Wilson loop matrices with respect to symmetries.

\begin{figure}
    \includegraphics[width=.8\columnwidth]{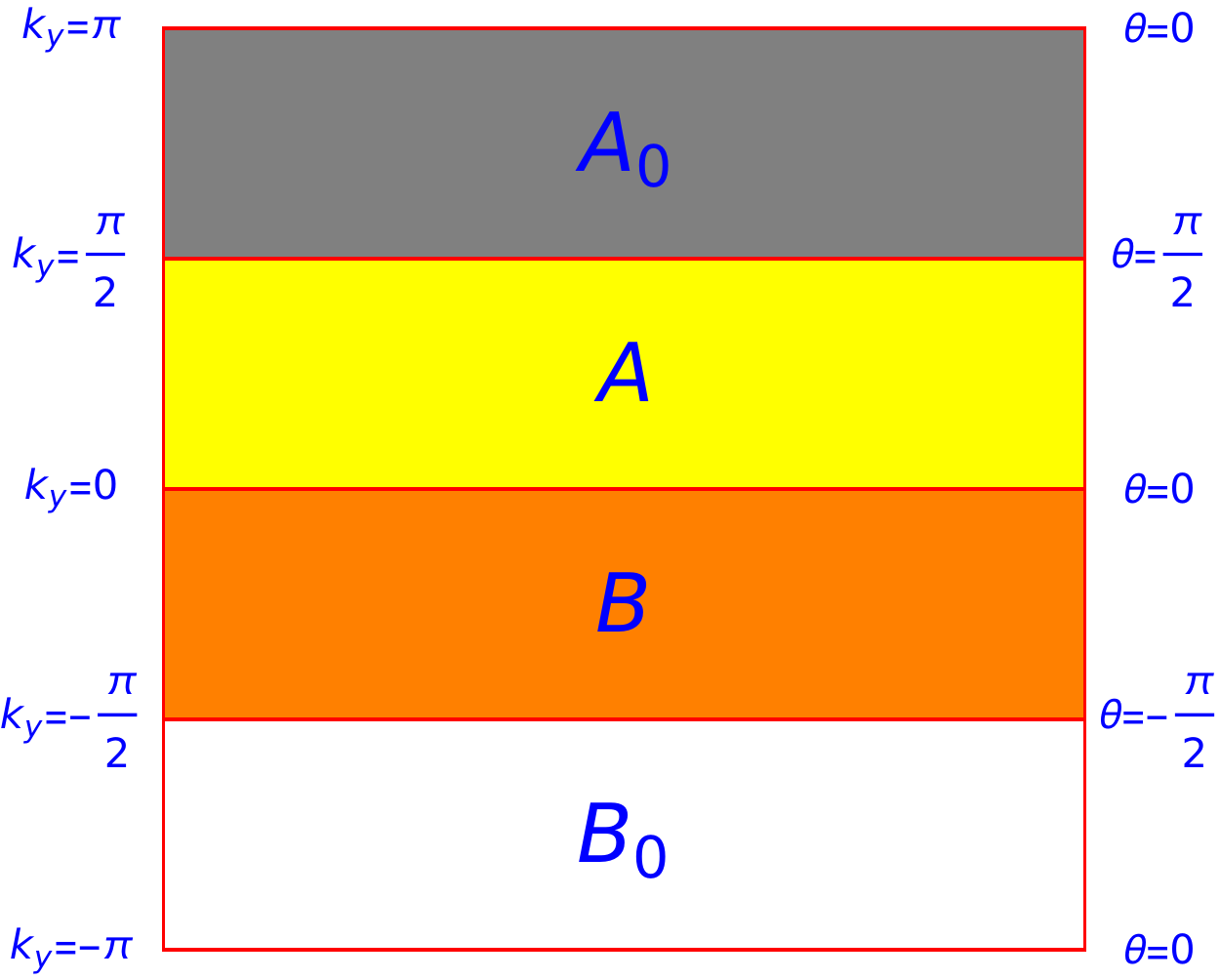}
    \caption[Patches]{Four patches in Brillouin zone. White color patch is denoted by $B_0$, orange color patch is denoted by $B$, yellow color patch is denoted by $A$, and gray color patch is denoted by $A_0$.} \label{fig:patches}
\end{figure}

We recall the Hamiltonian in the main text
\begin{align} \label{eq:hamAppd}
    H(k_x,k_y) = \cos \theta \begin{pmatrix}
        0 & q(k_x) \\
        q^T(k_x) & 0
    \end{pmatrix} + \sin \theta \begin{pmatrix}
        I_{4 \times 4} & O\\
        O & -I_{4 \times 4}
    \end{pmatrix},
\end{align}
where $\theta \in [-\pi/2,\pi/2]$, and $q(k_x)$ in the expression equals $q_1(k_x)$ for $k_y \in [-\frac{\pi}{2},\frac{\pi}{2}]$ and equals $q_0(k_x)$ for $k_y \in [-\pi,-\frac{\pi}{2}] \cup [\frac{\pi}{2},\pi]$. We divide the Brillouin zone into four patches, as shown in Fig. \ref{fig:patches}. On each patch, the periodic part of the occupied bands wave function can be analytically solved.
On patch $A$,
\begin{align}
    \ket{u^{\alpha}(k_x,k_y)} = \frac{1}{\sqrt{2+2 \sin \theta}} 
    \begin{pmatrix}
        \cos\theta \cdot q_1(k_x) \ket{v^\alpha} \\
        -(1+\sin \theta) \ket{v^\alpha}
    \end{pmatrix}.
\end{align}
On patch $B$,
\begin{align}
    \ket{u^{\alpha}(k_x,k_y)} = \frac{1}{\sqrt{2-2 \sin \theta}} 
    \begin{pmatrix}
        -(1-\sin \theta) \ket{v^\alpha} \\
        \cos \theta \cdot q_1^T(k_x) \ket{v^\alpha}
    \end{pmatrix}.
\end{align}
On patch $A_0$,
\begin{align}
    \ket{u^{\alpha}(k_x,k_y)} = \frac{1}{\sqrt{2+2 \sin \theta}} 
    \begin{pmatrix}
        \cos\theta \cdot q_0(k_x) \ket{v^\alpha} \\
        -(1+\sin \theta) \ket{v^\alpha}
    \end{pmatrix}.
\end{align}
On patch $B_0$,
\begin{align}
    \ket{u^{\alpha}(k_x,k_y)} = \frac{1}{\sqrt{2-2 \sin \theta}} 
    \begin{pmatrix}
        -(1-\sin \theta) \ket{v^\alpha} \\
        \cos \theta \cdot q_0^T(k_x) \ket{v^\alpha}
    \end{pmatrix}.
\end{align}
The $\ket{v^\alpha}, \alpha \in \{1,2,3,4\}$ in the above equations is the canonical orthonormal basis of $\mathbb{C}^4$,
\begin{align}
    \ket{v^1} = 
    \begin{pmatrix}
        1 \\ 0 \\ 0 \\0
    \end{pmatrix},
    \ket{v^2} =
    \begin{pmatrix}
        0 \\ 1 \\ 0 \\ 0
    \end{pmatrix},
    \ket{v^3} =
    \begin{pmatrix}
        0 \\ 0 \\ 1 \\ 0
    \end{pmatrix},
    \ket{v^4} =
    \begin{pmatrix}
        0 \\ 0 \\ 0 \\ 1
    \end{pmatrix}.
\end{align}

We show that this basis is parallel transported along $k_y$ on each patch, in other word, the Berry connection vanishes along $k_y$. On patch $A$, the elements of the Berry connection are expressed as
\begin{align}
    & \mathcal{A}^{\alpha \beta}(k_x,k_y) \notag \\
    = & \bra{u^\alpha (k_x,k_y)} d \ket{u^\beta(k_x,k_y)} \notag \\
    = & \frac{1}{\sqrt{2+2\sin \theta}} \begin{pmatrix}
        \bra{v^\alpha} q_1^T(k_x) \cos \theta & -\bra{v^\alpha} (1+\sin \theta)
    \end{pmatrix} \bigg [ \notag \\
    & -\frac{\cos \theta}{(2+2 \sin \theta)^{\frac{3}{2}}} \begin{pmatrix}
        \cos \theta \cdot q_1(k_x) \ket{v^\beta} \\
        -(1+\sin \theta) \ket{v^\beta}
    \end{pmatrix} \notag \\
& + \frac{1}{\sqrt{2+2\sin \theta}} \begin{pmatrix}
    -\sin \theta \cdot q_1(k_x) \ket{v^\beta} \\
    -\cos \theta \ket{v^\beta}
\end{pmatrix} \notag \\
& + \frac{1}{\sqrt{2+2\sin \theta}} \begin{pmatrix}
    \cos \theta \cdot d q_1(k_x) \ket{v^\beta} \\
    0
\end{pmatrix} \bigg ] \notag \\
    = & - \frac{\cos \theta}{2+2 \sin \theta} \delta^{\alpha \beta} +  \frac{\cos \theta}{2+2 \sin \theta} \delta^{\alpha \beta} + \notag \\
    & \frac{\cos^2 \theta}{2+2 \sin \theta} \bra{v^\alpha} q^T_1(k_x) \cdot d q_1(k_x) \ket{v^\beta} \notag \\
    = & \frac{\cos^2 \theta}{2+2 \sin \theta} \bra{v^\alpha} q^T_1(k_x) \cdot d q_1(k_x) \ket{v^\beta},
\end{align}
and in matrix form,
\begin{align} \label{eq:connection}
    \mathcal{A}(k_x,k_y) = \frac{\cos^2 \theta}{2+2 \sin \theta} q^T_1(k_x) \cdot d q_1(k_x).
\end{align}
Thus the Berry connection vanishes along the $\theta$($k_y$)-direction(there is no $d\theta$ term in Eq. (\ref{eq:connection})). This conclusion holds on all patches.

The transition function on the intersection of two patches can be computed as well.
\begin{align}
    t^{AB}(k_x)^{\alpha \beta}= \braket{u^{\alpha}_A(k_x,0)| u^{\beta}_B(k_x,0)}=-\bra{v^\alpha} q_1^T(k_x) \ket{v^\beta},
\end{align}
which has the matrix form 
\begin{align}
    t^{AB}(k_x) = -q_1^T(k_x).
\end{align}
Similarly, all the other transition functions are
\begin{align}
    t^{A_0 B_0}(k_x) & = -q_0^T(k_x) \notag \\
    t^{A_0 A}(k_x) & = I_{4 \times 4} \notag \\
    t^{B B_0}(k_x) & = I_{4 \times 4}.
\end{align}

The Wilson loop matrix(holonomy) is defined by an integral of parallel transports on patches and transition functions. Let $\{ U_i \}_{i=1,\dots,N}$ be a cover including the loop $l$. Divide $l$ to N pieces so that $l_i \subseteq U_i$. Let $p_i$ be the junction points of $l_i$, namely, $\partial l_i = p_{i+1} - p_i$. The Wilson loop is defined by \cite{PhysRevB.95.235425,PhysRevLett.121.106403}
\begin{align}
    W_{l} = t^{1,N}(p_1) \cdot P e^{-\int_{l_N} \mathcal{A}_N} \cdot t^{N,N-1}(p_N) \dots P e^{-\int_{l_1} \mathcal{A}_1},
\end{align}
where $\mathcal{A}_i$ is the Berry connection on $U_i$, and $t^{i,j}$ is the transition function on $U_i \cap U_j$.
In this example, since the connection vanishes along the $k_y$-direction, all $P e^{-\int_{l_i} \mathcal{A}_i}$ terms are equal to $I_{4 \times 4}$. Hence, the Wilson loop matrix (holonomy)is expressed by a formula only in terms of transition functions
\begin{align}
    W_{\mathbf{e}_2,\pi}(k_x) & = t^{A_0 A}(k_x) t^{A B}(k_x) t^{B B_0}(k_x) t^{B_0 A_0}(k_x) \notag \\
    & = q_1^T(k_x) q_0^T(k_x)^{-1},
\end{align}
which is the result used in the main text.

Furthermore, this process provides an explicit inverse map of the dimension raising map. In general, any $T,C_2$-symmetric Hamiltonian with a vanishing $\nu_{\mathbf{\Gamma X}}$ can be deformed to the form of Eq. (\ref{eq:hamAppd}), which suggests the following procedure. Given a Hamiltonian of a $T,C_2$-symmetric system, we calculate its Wilson loop matrix $W_{\mathbf{e}_2,\pi}(k_x)$, which corresponds to an one dimensional chiral system whose Hamiltonian is 
\begin{align}
    h_{\mathrm{chiral}}(k_x)=\begin{pmatrix}
        0 & W_{\mathbf{e}_2,\pi}(k_x) \\
        W^T_{\mathbf{e}_2,\pi}(k_x) & 0
    \end{pmatrix},
\end{align}
and the chiral symmetry is 
\begin{align}
    \Gamma = \begin{pmatrix}
        I_{4 \times 4} & 0 \\
        0 & -I_{4 \times 4}
    \end{pmatrix}.
\end{align}
This one dimensional chiral-symmetric system is still $T,C_2$-symmetric, and
\begin{align}
    T = \begin{pmatrix}
        0 & w \\
        w & 0
    \end{pmatrix},
    C_2 = \begin{pmatrix}
        0 & B \\
        B & 0
    \end{pmatrix},
\end{align}
where $w$ is the sewing matrix of TR-symmetry as in Eq. (\ref{eq: Wilson_T_symm}), and $B$ is the sewing matrix of twofold rotation symmetry. This fact implies that the topological equivalent class of the Hamiltonian with respect to symmetries is encoded by the topological equivalent class of the Wilson loop matrix with respect to symmetries (only the information of the partial polarization along the $x$-direction is lost, and it is encoded by the Wilson loop along another direction). In the following, we show that a homotopy of a Wilson loop with respect to symmetries gives a deformation of the Hamiltonian without breaking any symmetries.

Assume that we have found a homotopy between two paths of Wilson loop matrices with respect to symmetries, that is, a map $W : [0,2 \pi] \times [0,1] \rightarrow SO(N)$ such that
\begin{align}
    w W(k,t) w^{-1} = W^{-1}(-k,t),
    W(k,t=0) = W_{\mathbf{e}_2,\pi}(k).
\end{align}
We keep $q_0(k_x,t)$ invariant during the deformation, and $W(k,t)$ and $q_1(k_x,t)$ are related by 
\begin{align}
    W(k,t) = q_1^T(k,t) q_0^T(k)^{-1},
\end{align}
which in other word says,
\begin{align}
    q_1^T(k,t) = W(k,t) q_0^T(k).
\end{align}
Use the expression of the time-reversal operator $T = 
\begin{pmatrix}
    0 & -\sigma_z \otimes \sigma_0\\
    \sigma_z \otimes \sigma_0 & 0
\end{pmatrix} \mathcal{K}$. The TR-invariance of the Hamiltonian (\ref{eq:ham}) requires
\begin{align}
    (\sigma_z \otimes \sigma_0) q^T(k,t) = -q(-k,t) (\sigma_z \otimes \sigma_0).
\end{align}
We check this requirement on $q_1(k,t)$,
\begin{align}
    & (\sigma_z \otimes \sigma_0) q_1(k,t)^T (\sigma_z \otimes \sigma_0) \notag \\
    = & (\sigma_z \otimes \sigma_0) W(k,t) q_0(k)^T (\sigma_z \otimes \sigma_0) \notag \\
    = & (\sigma_z \otimes \sigma_0) W(k,t) (\sigma_z \otimes \sigma_0) 
    (\sigma_z \otimes \sigma_0) q_0(k)^T (\sigma_z \otimes \sigma_0) \notag \\
    = & (\sigma_z \otimes \sigma_0) W(k,t) (\sigma_z \otimes \sigma_0) (-q_0(-k)) \notag \\
    = & - (\sigma_z \otimes \sigma_0) W(k,t) w^{-1} \notag \\
    = & - (\sigma_z \otimes \sigma_0) w^{-1} W^T(-k,t) \notag \\
    = & - (\sigma_z \otimes \sigma_0) (\sigma_z \otimes \sigma_0) q_0 W^T(-k,t) \notag \\
    = & - q_0 W^T(-k,t) \notag \\
    = & - q_1(-k,t),
\end{align}
where in the fourth and the sixth equality the relation $w = q_0(k) (\sigma_z \otimes \sigma_0)= q_0 (\sigma_z \otimes \sigma_0)$ has been used.

Hence the problem of the classification of the Hamiltonian with respect to symmetries is equivalent to the problem of the classification of the Wilson loops along the $x$- and $y$-directions with respect to symmetries. For this reason, we focus on the classification of the Wilson loop matrices in the main text, as the homotopy between two paths of Wilson loop matrices induces a deformation of Hamiltonian.

\section{Time reversal related channels and the sewing matrix} \label{sec: sewing}
The sewing matrix of the TR-symmetry of the example in the main text is
\begin{align}
    & w^{\alpha \beta}(k_x,k_y=\pi) \notag \\
    = & \bra{u^{\alpha}(-k_x,\pi)} T \ket{u^{\beta}(k_x,\pi)} \notag \\
    = & \frac{1}{2} \begin{pmatrix}
        \bra{v^\alpha} q_0^T(k_x) & -\bra{v^\alpha} 
    \end{pmatrix} \cdot
    \begin{pmatrix}
        0 & -\sigma_z \otimes \sigma_0 \\
        \sigma_z \otimes \sigma_0 & 0
    \end{pmatrix} \notag \\
    & \cdot \begin{pmatrix}
        q_0(k_x) \ket{v^\beta} \\
        -\ket{v^\beta}
    \end{pmatrix} \notag \\
    = & \bra{v^\alpha} q_0(k_x) (\sigma_z \otimes \sigma_0) \ket{v^\beta},
\end{align}
which in matrix form is 
\begin{align}
    w(k_x,k_y=\pi) = q_0(k_x) (\sigma_z \otimes \sigma_0) = q_0 \cdot (\sigma_z \otimes \sigma_0).
\end{align}
Hence, the sewing matrix at the $k_y=\pi$ line is independent of $k_x$.

Next, we show that the subbundle of occupied bands can be divided into two channels, which are themselves $T C_2$-symmetric and are $T$-related to each other. Let $E_\mathrm{I}$($E_\mathrm{II}$) denote the channel I(II) subbundle.
\begin{align}
    E_\mathrm{I}= \bigsqcup_{k \in \mathbb{T}^2} \{
        \begin{pmatrix}
            0 \\ 0 \\ a \\ b \\ c \\ d \\ 0 \\ 0
        \end{pmatrix} 
        | a,b,c,d \in \mathbb{C} \} \cap E_{occ,k}, \notag \\
    E_{\mathrm{II}}= \bigsqcup_{k \in \mathbb{T}^2} \{
        \begin{pmatrix}
            a \\ b \\ 0 \\ 0 \\ 0 \\ 0 \\ c \\ d
        \end{pmatrix}
        | a,b,c,d \in \mathbb{C} \} \cap E_{occ,k},
\end{align}
where $E_{occ,k}$ is the fiber of the occupied bundle at momentum $k$. It is easy to see that these two channels are themselves $T C_2$-symmetric and are $T$-related to each other, that is, 
\begin{align}
    TC_2 E_\mathrm{I,II} & \subseteq E_\mathrm{I,II}, \notag \\
    T E_\mathrm{I} & \subseteq E_\mathrm{II}, \notag \\
    T E_\mathrm{II} & \subseteq E_\mathrm{I}.
\end{align}
They are indeed vector bundles, since if we consider a particular local patch, say patch $A$,
\begin{align}
    E_\mathrm{I}|_A = \bigsqcup_{k \in A} span_\mathbb{C} \{ \ket{u^1(k)},\ket{u^2(k)} \}, \notag \\
    E_\mathrm{II}|_A = \bigsqcup_{k \in A} span_\mathbb{C} \{ \ket{u^3(k)},\ket{u^4(k)} \},
\end{align}
where $u^\alpha(k),\alpha=1,2,3,4$ are four occupied bands we solved in appendix \ref{sec: Wilson}.

\section{Explicit expression of a symmetry preserving homotopy between two Wilson loops}
We construct in this appendix an explicit expression of a symmetry preserving homotopy between two Wilson loops. Wilson loops considered here are in the form of Eq. (\ref{eq: wil_example2_0}). In the main text, we stated that the Wilson loop with $m=1,n=3$ is homotopic to the Wilson loop with $m=2,n=4$. We have obtained an explicit expression of a homotopy between these two Wilson loops; however, it is quite sophisticated and we believe it is not suitable as the first demonstration example. Here, we present a simpler example, i.e., we construct an explicit expression of a homotopy between the Wilson loop with $m=1,n=1$ and the Wilson loop with $m=0,n=0$ (these two phases are all trivial, i.e. they have $\nu_{new}=0 \mod 2$). The same strategy is still valid for more sophisticated examples. 

\begin{figure}
    \centering
    \subfigure[]{
        \centering
        \includegraphics[width=.8\columnwidth]{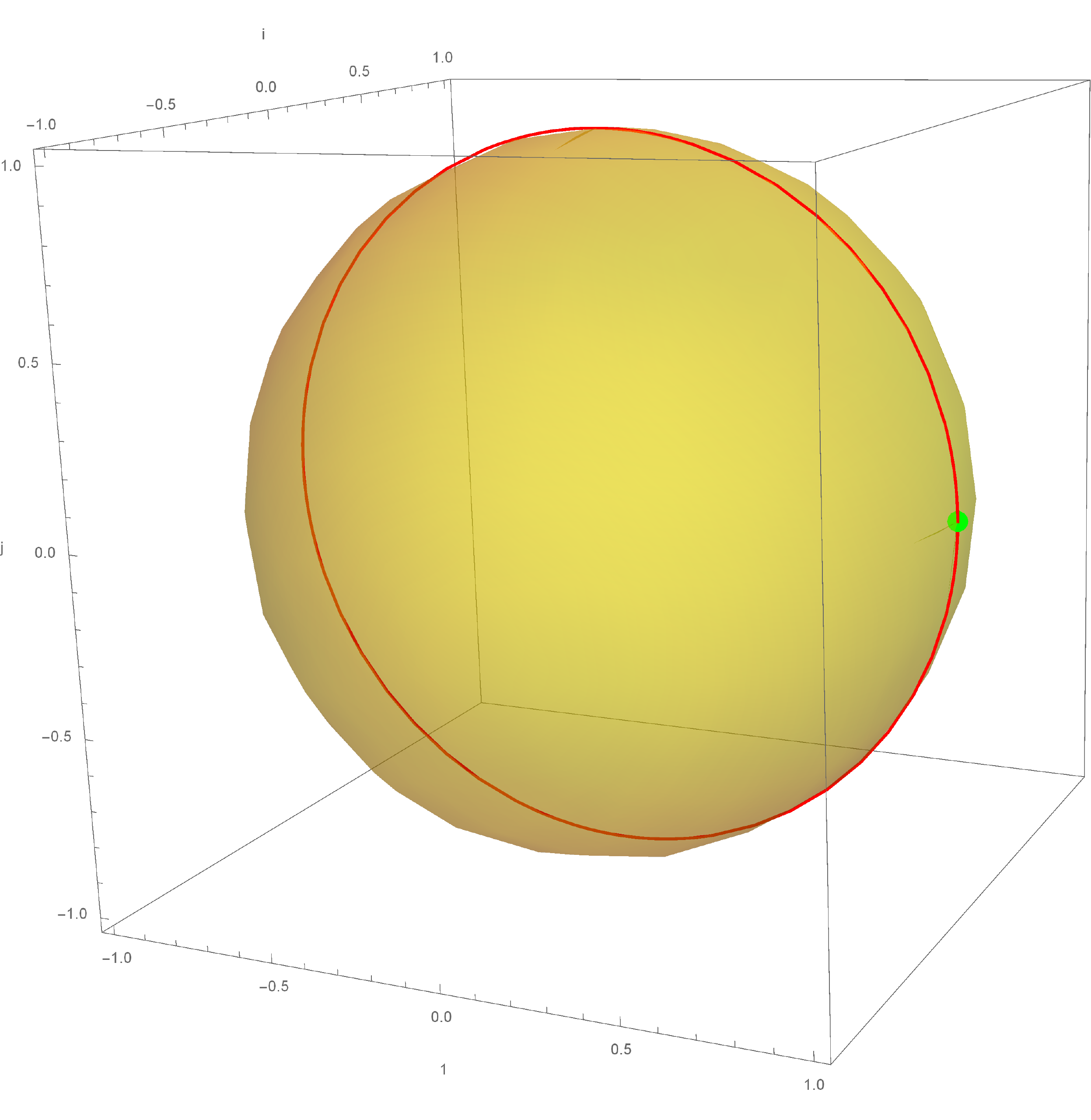}
    } \\
    \subfigure[]{
        \centering
        \includegraphics[width=.8\columnwidth]{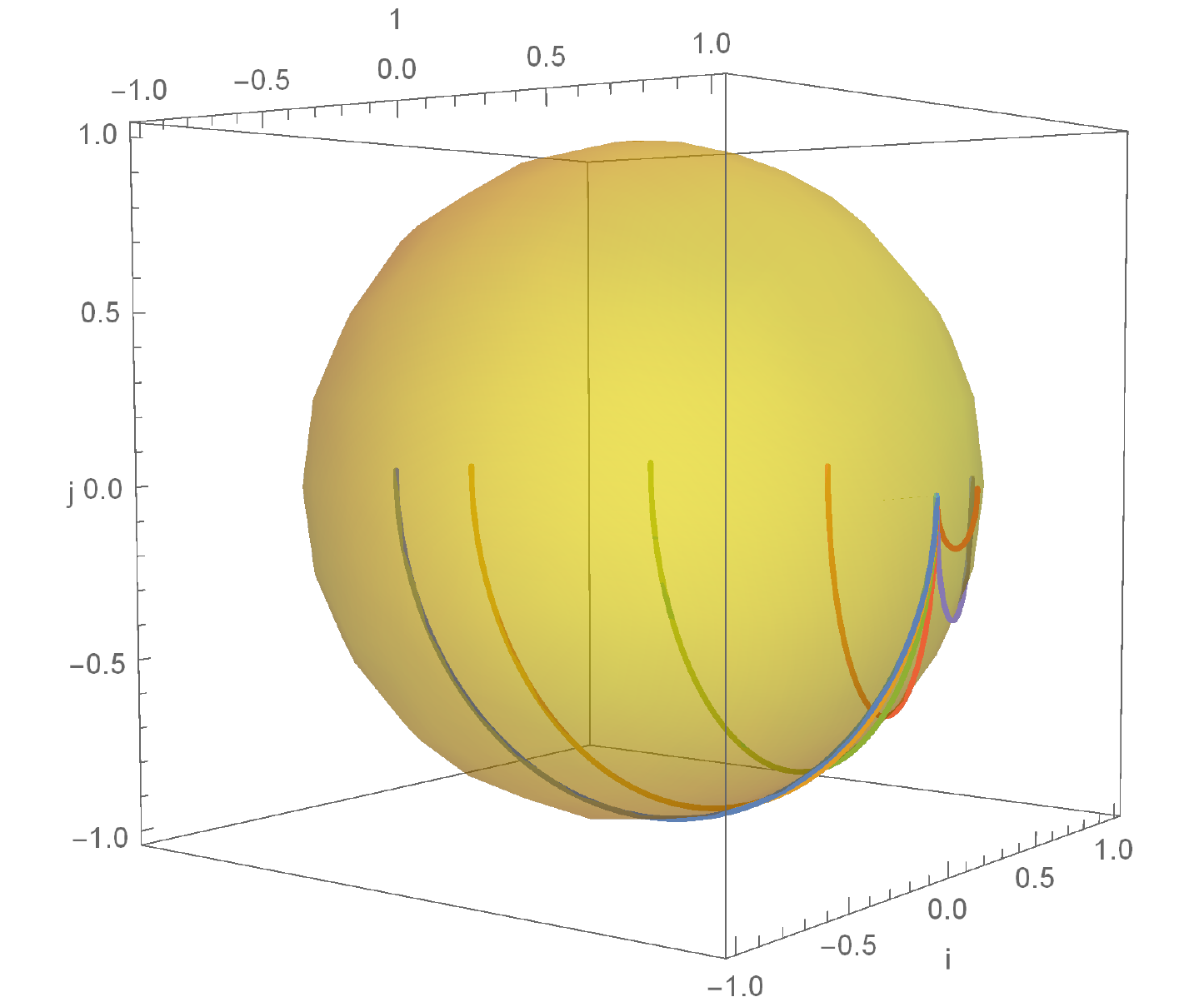}
    }
    \caption[The deformation of the second component of the lifted Wilson loop.]{The deformation of the second component of the lifted Wilson loop. (a)The initial (red) and final (green) lifted Wilson loop. (b)A series of deformation time slices, i.e. $pr_2(\tilde{W}(k_x,t))$ where $k_x \in [0,\pi]$ and $t=0,0.1,0.3,0.5,0.75,0.9,1$.} \label{fig: TwoLoops}
\end{figure}

The strategy of finding a symmetry preserving homotopy between two Wilson loops is first to find the symmetry preserving homotopy between their lifts, and then to project the homotopy back to the $SO(4)$ group via the covering map in Eq. (\ref{eq: covering_map}). Let $W_0(k_x)$ be the Wilson loop with $m=n=1$, and let $W_1(k_x)$ be the Wilson loop with $m=n=0$. By Eq. (\ref{eq: lifted_wil2}), their lifts are $\tilde{W}_0(k_x) = (1, \cos k_x - j \sin k_x)$ and $\tilde{W}_1(k_x) = (1,1)$ respectively. Since their first components are same, we keep the first component invariant during the deformation. Plot their second components in the three dimensional space spanned by $1,i,j \in \mathbb{H}$ as shown in Fig \ref{fig: TwoLoops}(a). The first loop is a large circle on a two dimensional sphere, and the second loop is a point(constant loop). We consider first the deformation of the first half of the loop (i.e. $k_x \in [0, \pi]$), and then map to the deformation of the other half via TR symmetry as the same strategy we applied in the proof of theorem \ref{thm: trivial}. The deformation process is shown in Fig \ref{fig: TwoLoops}(b). Note that the midpoint is always in the fixed point set $X_0$.
It is easy to calculate the second component of the homotopy as
\begin{align}
    & pr_2(\tilde{W}(k_x,t)) \notag \\
    = & [1-(1-\cos k_x) \sin^2\frac{\pi}{2}(1-t)] \notag \\
     & + (1-\cos k_x) \sin \frac{\pi}{2}(1-t) \cos \frac{\pi}{2}(1-t) \; i \notag \\
    & - \sin \frac{\pi}{2}(1-t) \sin k_x \; j , \; \mathbf{for} \; k_x \in [0,\pi] . \notag 
\end{align}
The deformation of the other half can be obtained via TR symmetry
\begin{align}
    \tilde{W}(k_x,t) & = \overline{\tilde{w}^{-1} \tilde{W}(2 \pi-k_x,t) \tilde{w}} \notag \\
    & = \overline{(1,-j) \tilde{W}(2 \pi-k_x,t) (1,j)} \notag \\
    & = (1, \, [1-(1-\cos k_x) \sin^2\frac{\pi}{2}(1-t)] \notag \\
    & + (1-\cos k_x) \sin \frac{\pi}{2}(1-t) \cos \frac{\pi}{2}(1-t) \; i \notag \\
   & - \sin \frac{\pi}{2}(1-t) \sin k_x \; j ) , \; \mathbf{for} \; k_x \in [\pi, 2\pi] . \notag 
\end{align}
Actually, the deformation of the lifted Wilson loop in $k \in [0,\pi]$ and $k \in [\pi,2\pi]$ has the same form, i.e.
\begin{align}
    \tilde{W}(k_x,t) & = (1, \, [1-(1-\cos k_x) \sin^2\frac{\pi}{2}(1-t)] \notag \\
    & + (1-\cos k_x) \sin \frac{\pi}{2}(1-t) \cos \frac{\pi}{2}(1-t) \; i \\
   & - \sin \frac{\pi}{2}(1-t) \sin k_x \; j ) , \; \mathbf{for} \; k_x \in [0, 2\pi] . \notag
\end{align}

\begin{figure}
    \centering
    \subfigure[]{
        \centering
        \includegraphics[width=.6\columnwidth]{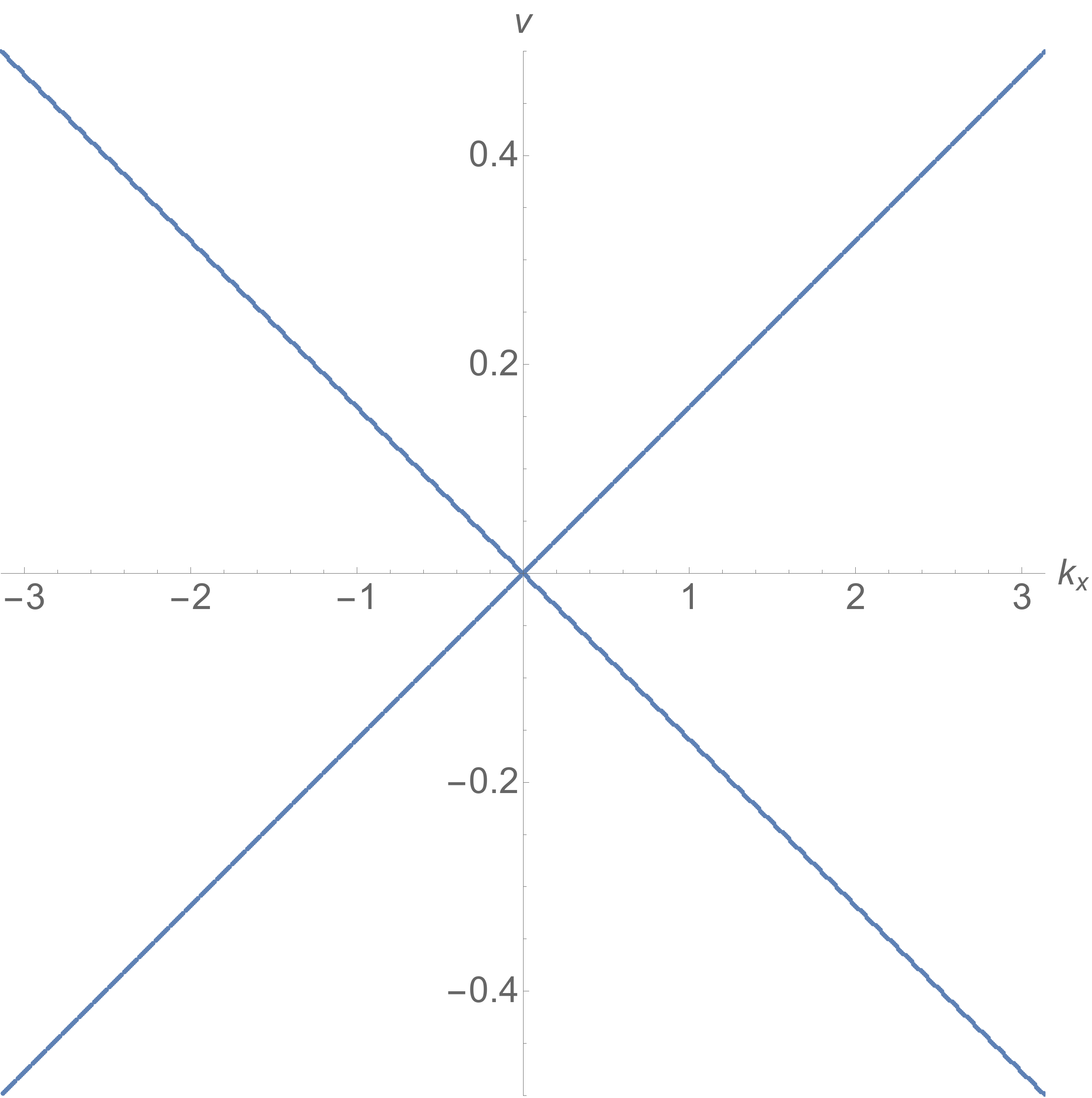}
    } \\
    \subfigure[]{
        \centering
        \includegraphics[width=.6\columnwidth]{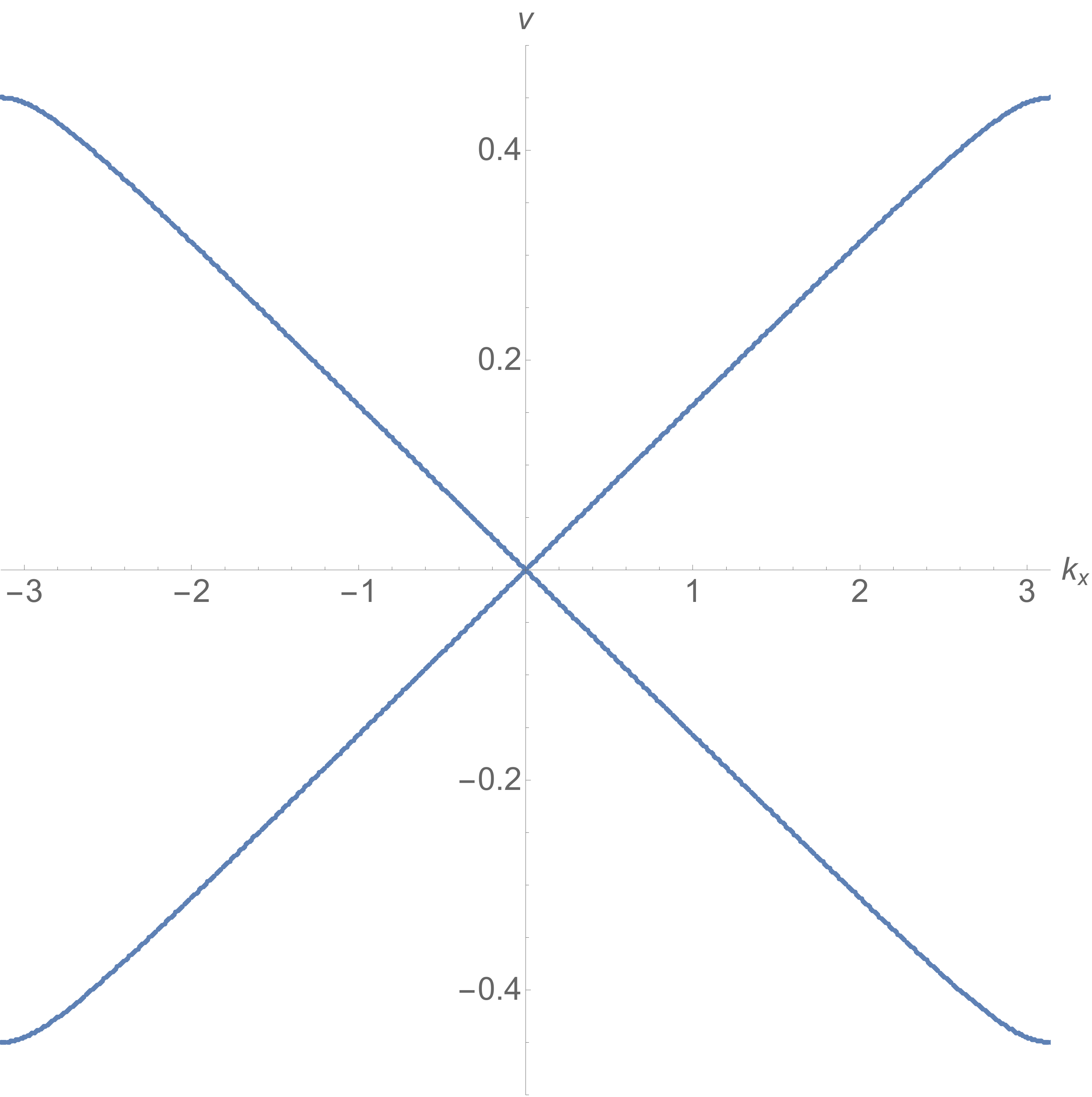}
    }
    \caption[The deformation of the Wilson loop spectrum.]{The deformation of the Wilson loop spectrum. (a)The Wilson loop spectrum of $W(k_x,t=0)$. It has non-zero winding numbers. (b)The Wilson loop spectrum of $W(k_x,t=0.1)$. It only has zero winding numbers.} \label{fig: wil_spectrum_deform}
\end{figure}

The deformation of the Wilson loop is obtained by projecting the deformation of the lifted Wilson loop to the $\text{SO}(4)$ group via the covering map in Eq. (\ref{eq: covering_map}), which is 
\begin{widetext}
    \begin{align}
        & W(k_x,t) = \notag \\
        & \left(
\begin{array}{cccc}
 \left(\cos \left(k_x\right)-1\right) \cos ^2\left(\frac{\pi  t}{2}\right)+1 & -\sin (\pi  t) \sin ^2\left(\frac{k_x}{2}\right) & \cos \left(\frac{\pi  t}{2}\right) \sin \left(k_x\right) & 0 \\
 \sin (\pi  t) \sin ^2\left(\frac{k_x}{2}\right) & \left(\cos \left(k_x\right)-1\right) \cos ^2\left(\frac{\pi  t}{2}\right)+1 & 0 & \cos \left(\frac{\pi  t}{2}\right) \sin \left(k_x\right) \\
 -\cos \left(\frac{\pi  t}{2}\right) \sin \left(k_x\right) & 0 & \left(\cos \left(k_x\right)-1\right) \cos ^2\left(\frac{\pi  t}{2}\right)+1 & \sin (\pi  t) \sin ^2\left(\frac{k_x}{2}\right) \\
 0 & -\cos \left(\frac{\pi  t}{2}\right) \sin \left(k_x\right) & -\sin (\pi  t) \sin ^2\left(\frac{k_x}{2}\right) & \left(\cos \left(k_x\right)-1\right) \cos ^2\left(\frac{\pi  t}{2}\right)+1 \\
\end{array}
\right) \notag \\
    & \in SO(4). 
    \end{align}
\end{widetext}
One can check $W(k_x,t)$ preserves the TR-symmetry, the initial Wilson loop $W(k_x,0)$ belongs to the second category with $m=n=1$, and $W(k_x,1)$ is the final Wilson loop which equals identity matrix. The Wilson loop spectrum of $W(k_x,t)$ has a set of winding numbers $\{1,1\}$ at $t=0$, and has zero winding numbers at infinitesimal $t$. Hence the winding numbers of the Wilson loop spectrum suddenly change at $t=0$ which is shown in Fig \ref{fig: wil_spectrum_deform}.

\section{Decomposition of bands into two time-reversal related channels} 
    \label{sec: Decomposition}
    In this section we present a method of decomposition of occupied bands into two time-reversal related channels. Such an algorithm has already been investigated in \cite{PhysRevB.85.115415}. However, since in this article we apply a Wilson loop approach method, and we do not require each occupied bands to be smooth, an alternative method will be used instead. We first simply review the concept of the Wilson loop operator, and then we illustrate our method. We illustrate our decomposition method for the case of two occupied bands with only TR symmetry, and the general case can be treated similarly.

Following the definition in \cite{PhysRevB.86.115112}, a discrete version of the Wilson loop operator is
\begin{align}
    & (W_{\mathbf{k}_1 \mathbf{k}_2})_{m n} \notag \\
    = & \sum_{a,b \dots} \braket{u_m(\mathbf{k}_1) | u_a(\mathbf{k}'_1)} \braket{u_a(\mathbf{k}'_1) | u_b(\mathbf{k}'_2)} \bra{u_b(\mathbf{k}'_2)} \dots \ket{u_n(\mathbf{k}_2)} , 
\end{align}
where $\mathbf{k}'_1 , \mathbf{k}'_2 , \dots$ form a path connecting $\mathbf{k}_1$ and $\mathbf{k}_2$, and $m,n,a$ and $b$ are indices of occupied bands.

This is by definition an $N_{occ} \times N_{occ}$ matrix. We can also define an operator $\hat{W}_{\mathbf{k}_1 \mathbf{k}_2}$ (a $(N_{occ}+N_{unocc}) \times (N_{occ}+N_{unocc})$ matrix) acting on spin-orbital space associated with this matrix:
\begin{align}
    \hat{W}_{\mathbf{k}_1 \mathbf{k}_2} & = \sum_{i,j \in occ} (W_{\mathbf{k}_1 \mathbf{k}_2})_{i j} \ket{u_i(\mathbf{k}_1)} \bra{u_j(\mathbf{k}_2)} \notag \\
    & = \ket{u_i(\mathbf{k}_1)} \braket{u_i(\mathbf{k}_1) | u_a(\mathbf{k}'_1)} \dots \ket{u_j(\mathbf{k}_2)} \bra{u_j(\mathbf{k}_2)} \notag \\
    & = P \exp \left( i \int_{\mathbf{k}_2}^{\mathbf{k}_1} \hat{P}_\mathbf{k} \partial \hat{P}_\mathbf{k} \cdot d\mathbf{k} \right),
\end{align}
where $\hat{P}_\mathbf{k} = \sum_{i \in occ} \ket{u_i(\mathbf{k})} \bra{u_i(\mathbf{k})}$ is the projector onto the occupied subspace at $\mathbf{k}$, and $P$ in the front of $\exp$ means "path ordered". We further denote
\begin{equation}
    \hat{W}_{\mathbf{e}_2,k_y}(k_x) = \hat{W}_{(k_x,k_y+2\pi) , (k_x,k_y)} ,
\end{equation}
where the path between the start point $(k_x,k_y)$ and the end point $(k_x,k_y+2 \pi)$ is a straight line, and $\mathbf{e}_2$ means the path is along the $k_y$ direction. The eigenvalues of this operator restricted on occupied bands subspace have modulus 1, which can be written as $\{ e^{i v_j(k_x)} | j=1,2,\dots N_{occ} \}$ for each fixed $k_x$, and they are independent of $k_y$ \cite{PhysRevB.96.245115}. We denote $\{ v_j(k_x) | j=1,2,\dots N_{occ} \}_{k_x}$ as the Wilson loop spectrum. Furthermore, each $v_j(k_x)$ is a Wannier center (center of the Wannier function) of the system \cite{PhysRevB.96.245115}.

We illustrate our decomposition algorithm by a quantum spin Hall insulator model. This insulator has a corresponding Hamiltonian
\begin{align} \label{eq:QSHham}
    h(\mathbf{k}) = {} &  \sin (k_x) (\Gamma_{zx} + \Gamma_{xx}) + \sin (k_y) (\Gamma_{yx}+\Gamma_{0y}) \notag \\
    & + [2-m-\cos (k_x)-\cos (k_y)] \Gamma_{0z} ,
\end{align}
where $\Gamma_{ij} = \sigma_i \otimes \tau_j$, and $\sigma_i(\tau_i)$ are Pauli matrices corresponding to the spin (orbital) degrees of freedom. 
\begin{figure}
    \includegraphics[width=.8\columnwidth]{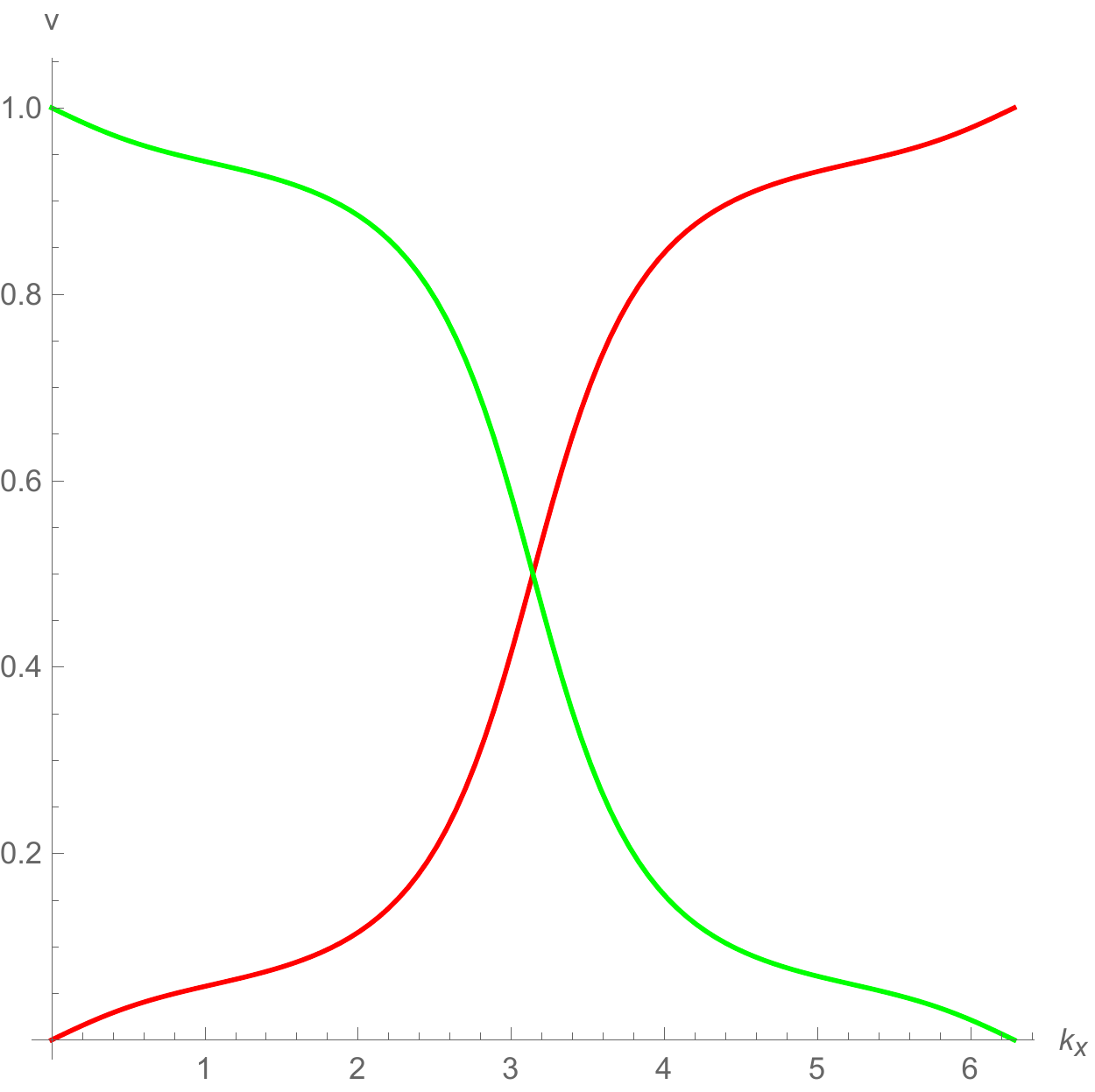}
    \caption[Wilson loop spectrum]{Wilson loop spectrum of Hamiltonian (\ref{eq:QSHham}) with $m=3$; two Wannier bands are plotted by different colors, i.e., red and green. The green Wannier band is labeled as band I, and the red Wannier band is labeled as band II.}\label{fig:QSHchannels}
\end{figure}
We plot its Wilson loop spectrum in Figure \ref{fig:QSHchannels}. In \cite{PhysRevB.96.245115}, the concept of Wannier bands is defined as the set of Wannier centers along x as a function of $k_y$, $v_x(k_y)$, or, vice versa, as the set of Wannier centers along y as a function of $k_x$, $v_y(k_x)$. We simply denote the green Wannier band and the red Wannier band in Figure \ref{fig:QSHchannels} by band I and band II, respectively. This system is time-reversal invariant, and the Wilson loop operator $W_{\mathbf{e}_2,k_y}(k_x)$ satisfies \cite{PhysRevB.96.245115}
\begin{equation} \label{eq:W_constraint}
    T \hat{W}_{\mathbf{e}_2,k_y}(k_x) T^{-1} = \hat{W}_{-\mathbf{e}_2,-k_y}(-k_x) = \hat{W}^{\dagger}_{\mathbf{e}_2,-k_y}(-k_x) ,
\end{equation}
where $T$ is the time-reversal operator, which is anti-unitary. Hence the set of Wannier centers satisfies the following constraint
\begin{equation}
    \{ v_j(k_x) \} \stackrel{T}{=} \{ v_j(-k_x) \},
\end{equation}
which in our case implies
\begin{equation} \label{eq:v_constraint}
    v_{\mathrm{I}}(k_x) = v_{\mathrm{II}}(-k_x) .
\end{equation}

In our case, the Wilson loop operator $\hat{W}_{\mathbf{e}_2,k_y}(k_x)$ at each $(k_x,k_y)$ has a spectral decomposition on the Hilbert space of the Hamiltonian,
\begin{align}
    \hat{W}_{\mathbf{e}_2,k_y}(k_x) = {} & e^{i 2\pi v_\mathrm{I}(k_x)}P_{\mathrm{I}}(k_x,k_y) + e^{i 2\pi v_{\mathrm{II}}(k_x)} P_{\mathrm{II}}(k_x,k_y) \notag \\
    & + 0 \cdot P_{\text{unocc}}(k_x,k_y) ,
\end{align}
where $P_\mathrm{I}(k_x,k_y)$ is the projection operator of $e^{i 2\pi v_\mathrm{I}(k_x)}$ eigenspace, $P_{\mathrm{II}}(k_x,k_y)$ is the projection operator of $e^{i 2\pi v_{\mathrm{II}}(k_x)}$ eigenspace, and $P_{\text{unocc}}(k_x,k_y)$ is the projection operator of the subspace of unoccupied bands (note that $P_\mathrm{I}$, $P_{\mathrm{II}}$, and $P_{\text{unocc}}$ are projectors on the eigenspace of the Hamiltonian). Hence,
\begin{align}
    & T \hat{W}_{\mathbf{e}_2,k_y}(k_x) T^{-1} \notag \\
    = & \; e^{-i 2\pi v_\mathrm{I}(k_x)} T P_{\mathrm{I}}(k_x,k_y) T^{-1} \notag \\
    & + e^{-i 2\pi v_{\mathrm{II}}(k_x)} T P_{\mathrm{II}}(k_x,k_y) T^{-1} \notag \\
    = & \; e^{-i 2\pi v_{\mathrm{II}}(-k_x)} T P_{\mathrm{I}}(k_x,k_y) T^{-1} \notag \\
    & + e^{-i 2\pi v_{\mathrm{I}}(-k_x)} T P_{\mathrm{II}}(k_x,k_y) T^{-1} ,
\end{align}
where in the second equality we have made use of (\ref{eq:v_constraint}). On the other hand,
\begin{align}
    & T \hat{W}_{\mathbf{e}_2,k_y}(k_x) T^{-1} = \hat{W}^{\dagger}_{\mathbf{e}_2,-k_y}(k_x)  \notag \\
    & = e^{-i 2\pi v_\mathrm{I}(-k_x)}P_{\mathrm{I}}(-k_x,-k_y) + e^{-i 2\pi v_{\mathrm{II}}(-k_x)} P_{\mathrm{II}}(-k_x,-k_y) .
\end{align}
Comparing the above two equations, a relation on projection operators can be obtained,
\begin{align}
    T P_{\mathrm{I}}(k_x,k_y) T^{-1} & = P_{\mathrm{II}}(-k_x,-k_y) \notag \\
    T P_{\mathrm{II}}(k_x,k_y) T^{-1} & = P_{\mathrm{I}}(-k_x,-k_y) ,
\end{align}
which means that two subbundles $Ran(P_{\mathrm{I}}(k_x,k_y))$ and $Ran(P_{\mathrm{II}}(k_x,k_y))$ are related by time-reversal symmetry.

\begin{figure} 
    \centering
    \subfigure[]{
        \centering
        \includegraphics[width=.4\columnwidth]{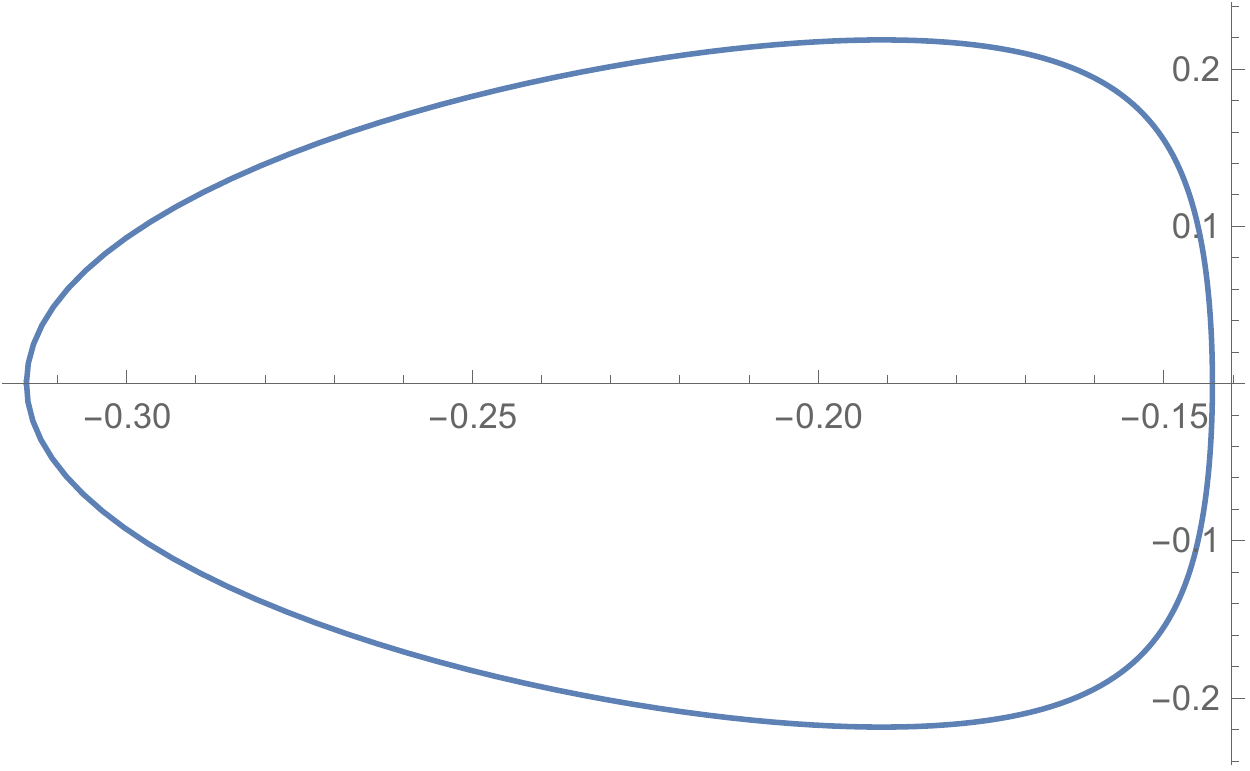}
    } 
    \subfigure[]{
        \centering
        \includegraphics[width=.4\columnwidth]{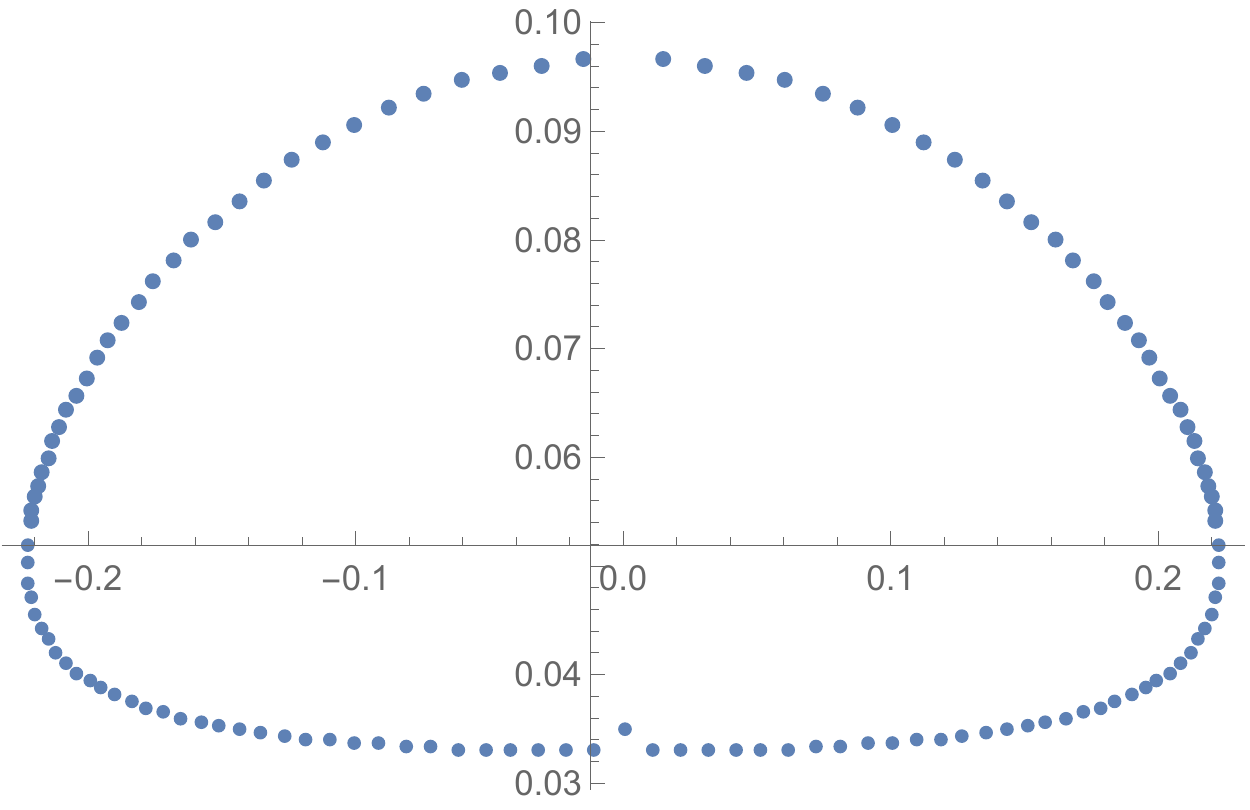}
    } 
    \caption[Continuity of the projection operator]{The entry $P_{\mathrm{I}}(k_x,k_y)_{1,2}$ along two circles in Brillouin zone. (a) The trajectory is chosen to be a straight line between $(0.3 \pi,-\pi)$ and $(0.3 \pi,\pi)$, which is actually a circle. (b) The trajectory is chosen to be a straight line between $(-\pi,0.2 \pi)$ and $(\pi, 0.2 \pi)$, which is also a circle. } \label{fig: Traj}
\end{figure}

We further show that the two projection operators $P_{\mathrm{I}}(k_x,k_y)$ and $P_{\mathrm{II}}(k_x,k_y)$ are continuous. Then $Ran(P_{\mathrm{I}}(k_x,k_y))$ and 
$Ran (P_{\mathrm{II}}(k_x,k_y))$ become two well-defined vector bundles. The two operators are continuous since the Wilson loop operator $W_{\mathbf{e}_2,k_y}(k_x)$ is continuous and two Wannier bands are chosen in a continuous way. We check this statement by numerically computing the trajectory of $P_{\mathrm{I}}(k_x,k_y)$ along some circles in the Brillouin zone which is shown in Figure \ref{fig: Traj}.

\bibliography{Main}
\end{document}